\documentclass[pra,twocolumn,superscriptaddress,aps,amsmath,amssymb,nofootinbib]{revtex4}
\usepackage{graphicx}
\usepackage{dcolumn}
\usepackage{bm}
\usepackage{color}
\usepackage[section]{placeins}
\usepackage{graphicx,epsfig}
\usepackage{amsmath}
\usepackage{amsfonts}
\usepackage{braket}
\usepackage{fancyhdr}
\usepackage{hyperref}
\usepackage[utf8]{inputenc}
\usepackage[T1]{fontenc}
\usepackage{amsthm}
\usepackage{hhline}
\usepackage{multirow}
\def\beq{\begin{equation}}
\def\eeq{\end{equation}}
\def\beqa{\begin{eqnarray}}
\def\eeqa{\end{eqnarray}}
\begin{document}
%
\title{Spin-orbit coupled bosons interacting in a two-dimensional harmonic trap}
\date{\today}
\author{Pere Mujal}
\affiliation{Departament de F\'{i}sica Qu\`{a}ntica i Astrof\'{i}sica,
Universitat de Barcelona, Mart\'{i} i Franqu\`{e}s 1, 08028 Barcelona, Spain}
\affiliation{Institut de Ci\`{e}ncies del Cosmos (ICCUB), Universitat 
de Barcelona, Mart\'{i} i Franqu\`{e}s 1, 08028 Barcelona, Spain}
\author{Artur Polls}
\affiliation{Departament de F\'{i}sica Qu\`{a}ntica i Astrof\'{i}sica,
Universitat de Barcelona, Mart\'{i} i Franqu\`{e}s 1, 08028 Barcelona, Spain}
\affiliation{Institut de Ci\`{e}ncies del Cosmos (ICCUB), Universitat de 
Barcelona, Mart\'{i} i Franqu\`{e}s 1, 08028 Barcelona, Spain}
\author{Bruno Juli\'{a}-D\'{i}az}
\affiliation{Departament de F\'{i}sica Qu\`{a}ntica i Astrof\'{i}sica,
Universitat de Barcelona, Mart\'{i} i Franqu\`{e}s 1, 08028 Barcelona, Spain}
\affiliation{Institut de Ci\`{e}ncies del Cosmos (ICCUB), Universitat 
de Barcelona, Mart\'{i} i Franqu\`{e}s 1, 08028 Barcelona, Spain}
\affiliation{Institut de Ci\`{e}ncies Fot\`{o}niques, Parc Mediterrani 
de la Tecnologia, 08860 Barcelona, Spain}
\begin{abstract}
A system of bosons in a two-dimensional harmonic trap in the presence of Rashba-type spin-orbit coupling is investigated. An analytic treatment of the ground state of a single atom in the weak-coupling regime is presented and used as a basis for a perturbation theory in the interacting two-boson system. The numerical diagonalization of both the single-particle and the two-boson Hamiltonian matrices allows us to go beyond those approximations and obtain not only the ground state, but also the low-energy spectra and the different energy contributions separately. We show that the expectation value of the spin-orbit term is related to the expectation value of $\hat{\sigma}_z \hat{L}_z$ for the eigenstates of the system, regardless of the trapping potential. The \textcolor{black}{low-energy states of the repulsively interacting two-boson system are characterized.} With the presence of a sufficiently strong interaction and spin-orbit coupling strength, there is \textcolor{black}{a direct} energy-level crossing in the ground state of the system \textcolor{black}{between states of different $J_z$, the third component of the total angular momentum,} that changes its structure. This is reflected in a discontinuity in the different energy terms and it is signaled in the spatial density of the system.
\end{abstract}
\maketitle
\section{Introduction}
Spin-orbit coupling in ultracold atoms~\cite{Dalibard,Zhang,Galitski,Goldman,Zhai,Machon} has been an issue of great interest in the last years in the atomic physics community. Since the first experiment was carried out successfully~\cite{Lin} dressing the atoms with two Raman lasers, additional investigations have been performed. For example, studying temperature effects~\cite{Ji} or engineering the spin-orbit coupling in alternative ways: with a gradient magnetic field~\cite{Luo}; and within optical lattices~\cite{Wu,Grusdt,Yamamoto}.
Interesting phenomena have been observed in spin-orbit coupled systems, for instance, a negative effective mass~\cite{Khamehchi}.

In the absence of a confining potential, in a homogeneous system, the single-particle energy dispersion relation is simple and the Hamiltonian is solvable in momentum space in the presence of spin-orbit coupling. In that case, for the many-body system at zero temperature, two phases were predicted in Ref.~\cite{Wang} in a mean-field approximation: the plane wave phase and the standing wave phase. The transition from one phase to the other was characterized depending on the inter- and intra-spin interactions between the atoms. Further studies in exploring the phase diagram of spin-orbit coupled Bose-Einstein condensates have been done within a mean-field description~\cite{Li2}, studying the stability of the system against quantum and thermal fluctuations~\cite{Baym,Ozawa2,Ozawa3,Ozawa4,Kawasaki}.

In the presence of a confining harmonic trap, the situation is fairly different, due to the introduction of a new characteristic length and the fact that the momentum is no longer a good quantum number. At the single-particle level, even when the spin-orbit coupling is strong, the spectrum remains discrete forming a Landau-level-like structure~\cite{Anderson,Zhou1,Sinah,Li,Hu}, which is altered when the trap is anisotropic~\cite{Marchukov,Marchukov2}. At the mean-field level, more phases, like a half-quantum vortex state, are found in the trapped system~\cite{Sinah,Hu,Ramachandhran,Zhou2,Zhou1,Li,Armaitis}.

The inclusion of interactions between the atoms adds an additional challenge, specially in the strongly interacting regime~\cite{Yin,Guan2,Schillaci,Guan}, where quantum correlations are expected to dominate the physics~\cite{Ramachandhran2}. Then, methods that go beyond mean field are required~\cite{Juan}.

In this work we make use of analytical approaches and numerical diagonalization techniques in order to describe the trapped single-particle and two-boson systems in the presence of Rashba spin-orbit coupling.

In Sec.~\ref{Sec.II}, the ground state of the single-particle system and the first low-energy states are computed and analyzed.
We relate the different energy contributions and also the expectation values of different kind of spin-orbit coupling terms applying the virial theorem.
In Sec.~\ref{Sec.III}, the interacting two-boson system is studied.
First, we give the second-quantized $N$-boson Hamiltonian and explain the methodology to diagonalize it for the $N=2$ case. In second place, we discuss the degeneracy breaking in the three-fold degenerate ground-state subspace.
In Sec.~\ref{Sec.IV}, we analyze the combined effects of the spin-orbit coupling and a spin-independent repulsive interaction in the spectrum. In particular, we find a crossover in the ground state characterized by a discontinuity in the energy contributions as a function of the spin-orbit \textcolor{black}{coupling strength} and by a change in the density profile of the system. Finally, conclusions and summary are presented in Sec.~\ref{Sec.V}.
\section{The single-particle system}
\label{Sec.II}
The physics of a particle of mass $m$ in a two-dimensional isotropic 
harmonic potential of frequency $\omega$ with Rashba type spin-orbit 
coupling is described by the Hamiltonian

\beq
\label{eq1ham}
\begin{split}
\hat{H}_{\rm sp} &= \frac{1}{2}m\omega^2\left(\hat{x}^2+\hat{y}^2\right)+\frac{\hat{p}^2_x+\hat{p}^2_y}{2m}+\frac{\kappa^2}{2m}
\\
&+\frac{\kappa}{m}\left ( \hat{\sigma}_x \hat{p}_x +\hat{\sigma}_y \hat{p}_y \right ),
\end{split}
\eeq
where $\kappa$ is the spin-orbit \textcolor{black}{coupling strength} and $\hat{\sigma}_x$ 
and $\hat{\sigma}_y$ are Pauli matrices. In the present paper, as we 
consider a bosonic system of ultracold atoms, the spin part does not 
refer to the intrinsic spin but to an internal degree of freedom or pseudospin, for instance, two hyperfine atomic states as in Ref.~\cite{Lin}. The Hamiltonian is composed by the kinetic energy, 
$\hat{K}=(\hat{p}^2_x+\hat{p}^2_y)/(2m)$, the harmonic potential, 
$\hat{V}_{\rm ho}=(m/2)\omega^2\left(\hat{x}^2+\hat{y}^2\right)$, 
the spin-orbit coupling, 
$\hat{V}_{\rm so}=(\kappa/m)\left ( \hat{\sigma}_x \hat{p}_x +\hat{\sigma}_y \hat{p}_y \right )$, and the constant term $\kappa^2/(2m)$. 
As mentioned in Ref.~\cite{Schillaci}, up to a pseudospin rotation, an alternative and equivalent form of the Rashba term would be 
$\propto \left ( \hat{\sigma}_x \hat{p}_y 
-\hat{\sigma}_y \hat{p}_x \right )$.

From now on, we use harmonic oscillator units, i.e., the energy is measured in units of $\hbar \omega$ and the length in units of 
\textcolor{black}{$x_{\rm ho}\equiv \sqrt{\hbar/(m \omega)}$}. The Hamiltonian in Eq.~(\ref{eq1ham}) is written in terms of annihilation operators, $\hat{a}_x=(\hat{x}+i\hat{p}_x)/\sqrt{2}$ and $\hat{a}_y=(\hat{y}+i\hat{p}_y)/\sqrt{2}$, and the corresponding creation operators, $\hat{a}^\dagger_x$ and $\hat{a}^\dagger_y$, as
\beq
\label{hamaadxy}
\begin{split}
\hat{H}_{\rm sp}&=(\hat{n}_x+\hat{n}_y +1)
\\
&+\frac{i\kappa}{\sqrt{2}} \left (\hat{\sigma}_x \left ( \hat{a}^\dagger_x-\hat{a}_x\right) +\hat{\sigma}_y \left (\hat{a}^\dagger_y-\hat{a}_y\right)\right )+\frac{\kappa^2}{2}\,.
\end{split}
\eeq
These operators fulfill the commutation relations 
$[\hat{{a}}_i,\hat{a}^\dagger_j]=\delta_{ij}$ and 
$[\hat{{a}}_i,\hat{a}_j]=[\hat{{a}}^\dagger_i,\hat{a}^\dagger_j]=0$, 
with $i,j=x,y$. We have used the number operators 
$\hat{n}_x= \hat{a}^\dagger_x\hat{a}_x$ and 
$\hat{n}_y= \hat{a}^\dagger_y\hat{a}_y$. Notice that $\kappa$ is not 
a dimensionless parameter in the original Hamiltonian, Eq.~(\ref{eq1ham}), and it is written in units of $\sqrt{\hbar m \omega}$ in Eq.~(\ref{hamaadxy}).

The single-particle basis can be labeled as, $\{\ket{n_x,n_y,m_s}\}$, 
with $n_x,n_y=0,1,2,\,...\,$, and $m_s=-1,1$, where $n_x$, $n_y$ and $m_s$ are eigenvalues of $\hat{n}_x$, $\hat{n}_y$ and $\hat{\sigma}_z$, respectively. 

\textcolor{black}{The matrix elements of the single-particle Hamiltonian written 
using the basis introduced above read
\beq
\label{ham0matrixelements0}
\bra{\alpha}\hat{H}_{\rm sp}\ket{\beta}=\epsilon_{\alpha,\beta}+\frac{\kappa^2}{2}\delta_{\alpha,\beta},
\eeq
with
\beqa
\epsilon_{\alpha,\beta}&=&\left(n_x(\alpha)+n_y(\alpha)+1\right)\delta_{\alpha,\beta} +\frac{i\kappa}{\sqrt{2}}\, \delta_{m_s(\alpha),-m_s(\beta)} \nonumber\\
&\times&
\Bigl(\sqrt{n_x(\beta)+1}\,\,\delta_{n_x(\alpha),n_x(\beta)+1}\,\delta_{n_y(\alpha),n_y(\beta)}\nonumber\\
&-& \sqrt{n_x(\beta)}\,\,\delta_{n_x(\alpha),n_x(\beta)-1}\,\delta_{n_y(\alpha),n_y(\beta)}\nonumber\\
&+&i\, m_s(\beta) \sqrt{n_y(\beta)+1}\,\,\delta_{n_x(\alpha),n_x(\beta)}\,\delta_{n_y(\alpha),n_y(\beta)+1}\nonumber\\
&-&i\, m_s(\beta) \sqrt{n_y(\beta)}\,\,\delta_{n_x(\alpha),n_x(\beta)}\,\delta_{n_y(\alpha),n_y(\beta)-1}
\Bigr)
\label{ham0matrixelements}
\eeqa
and $\ket{\alpha} \equiv \ket{n_x(\alpha),n_y(\alpha),m_s(\alpha)}$. The index $\alpha$ labels 
each state of the single-particle basis. The Hamiltonian matrix is fully diagonalized using the first $5112$ states in order of increasing energy 
$\epsilon_{\alpha,\alpha}$, which corresponds to $(n_x+n_y) \leqslant 70$ and $m_s=-1,1$.}
With this truncated Hilbert space the energies obtained are upper bounds 
to the exact ones. The method is variational, since we diagonalize in a subspace of the full Hilbert space.
\begin{figure}[t]
\centering
\includegraphics[width=\columnwidth]{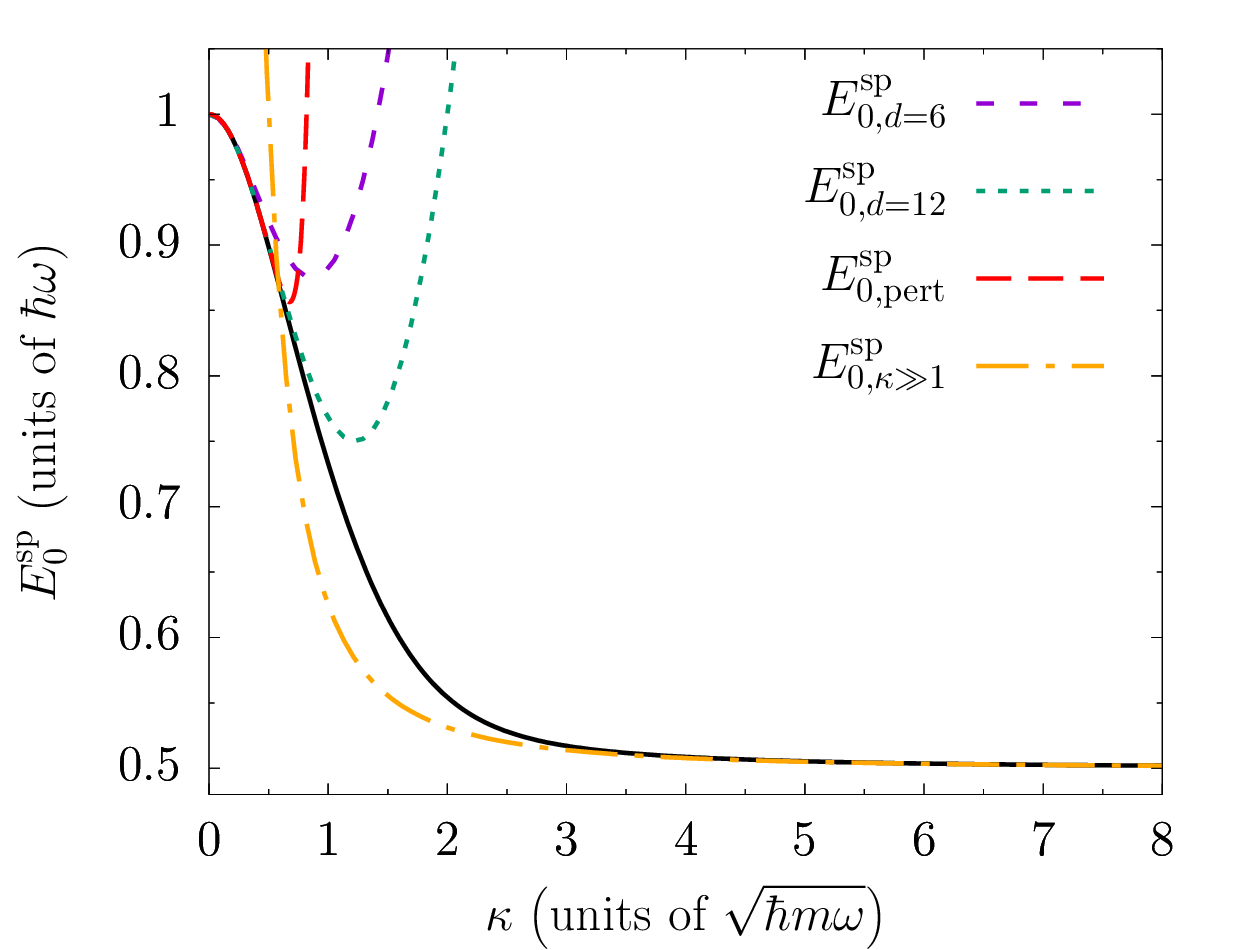}
\caption{\textcolor{black}{Solid black line: Single-particle ground-state energy, $E_0^{\rm sp}$, of the Hamiltonian in Eq.~(\ref{hamaadxy}) computed by numerical diagonalization. Medium-dashed purple line: $E_{0,d=6}^{\rm sp}$ given in Eq.~(\ref{eqtrunc6}). Short-dashed green line: $E_{0,d=12}^{\rm sp}$ given in Eq.~(\ref{eqtrunc12}). Long-dashed red line: Perturbative energy from Ref.~\cite{Yin}, $E_{0,{\rm pert}}^{\rm sp}$, given in Eq.~(\ref{eqgroundpert}). Dashed-dotted orange line: Limit value for $\kappa \gg 1$, $E_{0,\kappa \gg 1}^{\rm sp}$, from Ref.~\cite{Sinah}, given in Eq.~(\ref{eqE0kappabig}).}}
\label{Fig:1}
\end{figure}
\subsection{\textcolor{black}{The single-particle ground state}}
\label{Sec.IIA}
In this section, we explore the transition from the weak spin-orbit 
coupling regime, $\kappa < 1$, to the strong spin-orbit coupling 
one, $\kappa \gg 1$, at the single-particle level. Our direct diagonalization 
results are compared with previously derived analytical expressions valid
for the $\kappa \gg 1$ limit in Ref.~\cite{Sinah}, with perturbation theory 
expressions, $\kappa \ll 1$, derived in Ref.~\cite{Yin}, and with our own truncated analytic predictions valid in the $\kappa \lesssim 1$ regime. 

In Fig.~\ref{Fig:1} we report the single-particle ground-state energy 
as a function of $\kappa$. The ground state is in all cases two-fold degenerated. For $\kappa=0$, we recover the harmonic oscillator 
result, $E_0^{\rm sp}=1$. As $\kappa$ is increased, the ground-state energy 
decreases towards an almost constant value of $E_0^{\rm sp}\simeq 0.5$, 
which is already reached for $\kappa\simeq 3$.

For $\kappa < 1$, we derive analytical approximate expressions for 
the ground state of the single-particle Hamiltonian and its energy. 
The variational method consists in truncating the Hilbert space to a 
small number of modes (see Appendix~\ref{ap1} for details). Analytic 
expressions can be obtained truncating to six or twelve modes, 
\beqa
\label{eqtrunc6}
E_{0,d=6}^{\rm sp}&=& \frac{1}{2}\left(3-\sqrt{4\kappa^2+1}\right)+\frac{\kappa^2}{2} \,, \\
\label{eqtrunc12}
E_{0,d=12}^{\rm sp}&=&2-\sqrt{2\kappa^2+1}+\frac{\kappa^2}{2}.
\eeqa

The goodness of these expressions is shown in Fig.~\ref{Fig:1}, comparing them with the direct diagonalization and also with the perturbative calculations performed in Ref.~\cite{Yin}, that we write in our units as:
\beq
\label{eqgroundpert}
\begin{split}
E_{0,{\rm pert}}^{\rm sp}&=1-\frac{1}{2}\kappa^2+\frac{1}{2}\kappa^4-\frac{2}{3}\kappa^6+\frac{79}{72}\kappa^8
\\
&-\frac{274}{135}\kappa^{10}+\frac{130577}{32400}\kappa^{12}.
\end{split}
\eeq
Eq.~(\ref{eqtrunc12}) is seen to provide the best approximation to the 
direct diagonalization results, providing an accurate description up to 
$\kappa=1$. The perturbative expression of Ref.~\cite{Yin}, Eq.~(\ref{eqgroundpert}), reproduces well the results up to $\kappa\simeq 0.7$ while the approximation with six modes already fails for $\kappa \simeq 0.5$. \textcolor{black}{The truncated analytical expressions fail to describe the ground state when it has relevant contributions from basis states that are not in the truncated subspace considered.}

The large $\kappa$ domain has been studied previously in 
Refs.~\cite{Anderson,Zhou1,Sinah,Li,Hu}. In this regime, approximate 
expressions for the two-degenerate states that define the 
ground-state subspace are given in Ref.~\cite{Sinah}, together with 
an expression for the  ground-state energy, 
\beq
\label{eqE0kappabig}
E_{0,\kappa \gg 1}^{\rm sp}=\frac{1}{2}+\frac{1}{8\kappa^2}.
\eeq
This approximation is in very good agreement with our numerical 
results for $\kappa>2$ (see Fig.~\ref{Fig:1}). In particular, 
they correctly capture the limiting value in the spin-orbit dominated 
regime, $E_0^{\rm sp}\to 1/2$.
\subsection{The single-particle energy spectrum}
\label{Sec.IIB}
\begin{figure}[t!]
\centering
\includegraphics[width=\columnwidth]{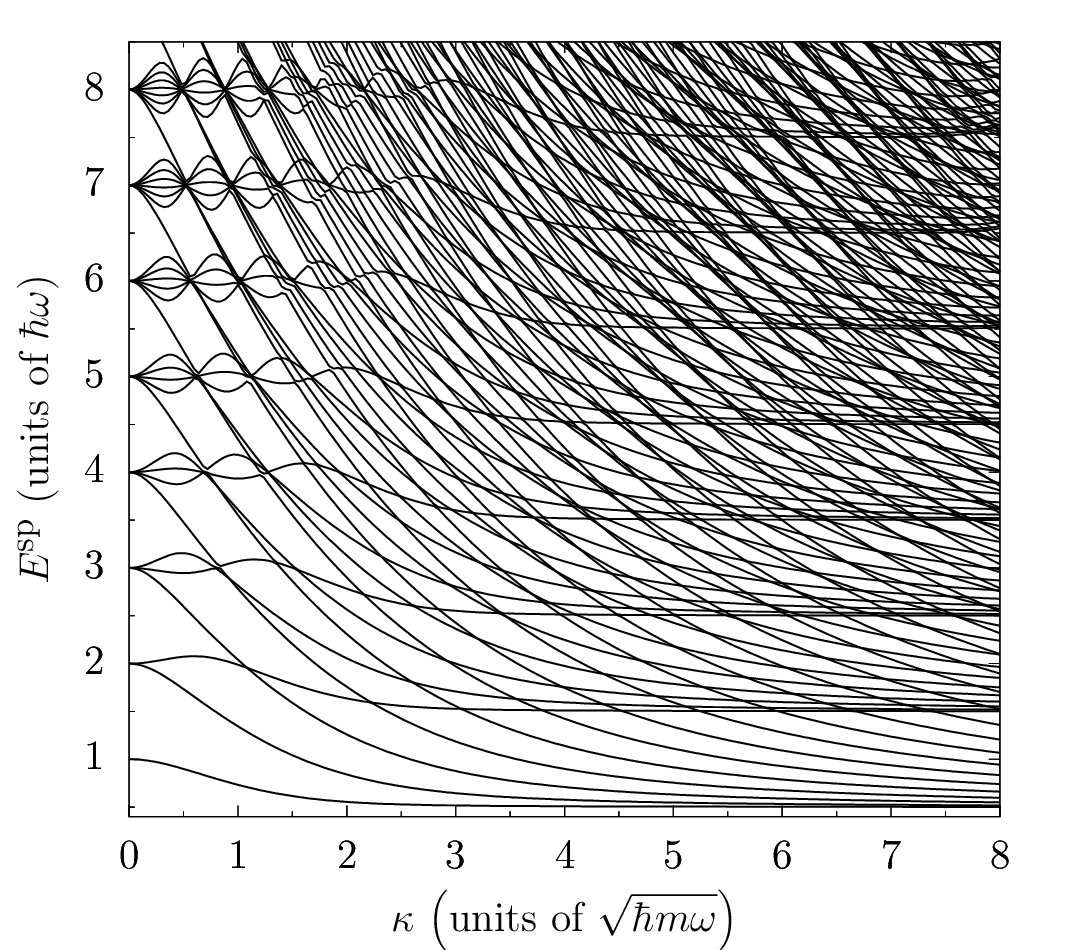}
\caption{\textcolor{black}{Energy spectrum of the single-particle Hamiltonian in Eq.~(\ref{hamaadxy}) depending on the spin-orbit \textcolor{black}{coupling strength}, $\kappa$. Notice that each energy is doubly degenerate so each line in the plot represents two equal energies that can be associated to two orthogonal eigenstates. This energy spectrum is also in the left panel of Fig.~1 of Ref.~\cite{Marchukov} up to $\kappa \approx 1$ and energies up to $20$, and in Fig.~3 of Ref.~\cite{Yin} for the lowest-energy eigenstates. The three-dimensional analogous spectrum is presented in Fig.~1 of Ref.~\cite{Anderson}.}}
\label{Fig:2}
\end{figure}

One of the important advantages of direct diagonalization methods 
is that they also provide, besides the \textcolor{black}{ground-state} properties, the low-energy part of the spectrum. The low-energy spectrum of the single-particle Hamiltonian, Eq.~(\ref{hamaadxy}), is depicted in Fig.~\ref{Fig:2}. 

In the limiting case of $\kappa=0$, the eigenstates of the Hamiltonian are the eigenstates of two independent two-dimensional harmonic oscillators, 
one for each spin component. Therefore, the energies are $E^{\rm sp}_n=n+1$ 
with degeneracy $2(n+1)$ and $n=n_x+n_y$. The case of $\kappa<1$ is 
analyzed in Ref.~\cite{Yin}, where the exact numerical values are 
compared with perturbation theory calculations in $\kappa$.

For any value of $\kappa$, all energy levels are two-fold 
Kramers-degenerate because the Hamiltonian is time-reversal 
symmetric~\cite{Ramachandhran,Marchukov,Hu,Zhou1,Li}. This degeneracy can be 
broken introducing a Zeeman term~\cite{Marchukov}. The effect 
of deforming the trap was considered in Ref.~\cite{Marchukov2}, 
which results in a breaking of the cylindrical symmetry of the 
system. In our case, the time-reversal symmetry is preserved and, 
in order to distinguish between the pair of degenerate states, 
we label them with $A$ and $B$, respectively, for a given energy $E^{\rm sp}$. The action of the time reversal operator, 
$\hat{T}=i\hat{\sigma}_y\mathcal{C}$~\cite{Ramachandhran,Marchukov,Hu,Li}, 
with $\mathcal{C}$ the complex conjugation operator, on the two-fold 
degenerate eigenstates reads\textcolor{black}{
\beqa
\ket{\psi^{\rm sp}_{E,B}}&=& i\hat{\sigma}_y\mathcal{C}\ket{\psi^{\rm sp}_{E,A}}\,, \nonumber \\
\ket{\psi^{\rm sp}_{E,A}}&=& i\hat{\sigma}_y\mathcal{C}\ket{\psi^{\rm sp}_{E,B}} \,.
\eeqa
}

The eigenstates of the single-particle Hamiltonian can be written 
in a basis with a well defined total angular momentum,
\beq
\label{eq1:AngularMomentum}
\hat{\bm{J}}=\hat{\bm{S}}+\hat{\bm{L}},
\eeq
where $\hat{\bm{S}}=(\hat{\sigma}_x,\hat{\sigma}_y,\hat{\sigma}_z)/2$, 
and $\hat{\bm{L}}\equiv\hat{\bm{r}}\times \hat{\bm{p}}=(0,0,\hat{L}_z)$, \textcolor{black}{both written in units of $\hbar$}.
The single-particle Hamiltonian commutes with $\hat{J}^2$ and $\hat{J}_z$.
Therefore, the eigenstates of the system can be labeled with the corresponding quantum numbers, $j$ and $j_z$, respectively, 
regardless of the value of $\kappa$. In particular, in the limiting case 
$\kappa \gg 1$, an additional radial quantum number, $n_r$, is 
introduced to describe the eigenstates of the system (see Ref.~\cite{Sinah}) and also the eigenenergies, approximately,
\beq
\label{Eq.energiesapprox}
E_{\kappa \gg 1}^{\rm sp} = n_r + \frac{1}{2}+\frac{j^2_z}{2\kappa^2},
\eeq
with $n_r=0,1,\,...\,$, and $j_z=m_l+1/2$, with $m_l=0,\pm 1,\,...\,$. 
The two-fold degeneracy is reflected in the fact that the energy 
depends on $j^2_z$, so it is independent of its sign. The eigenstates 
with the same radial quantum number, $n_r$, tend to become degenerate 
with increasing $\kappa$, forming an energy manifold. This kind of 
physics has been studied in two and three dimensions, where the same 
type of Landau-level-like spectrum is found and described in terms of dimensional reduction~\cite{Anderson,Zhou1,Sinah,Li,Hu}.
\begin{figure}[t!]
\centering
\includegraphics[width=\columnwidth]{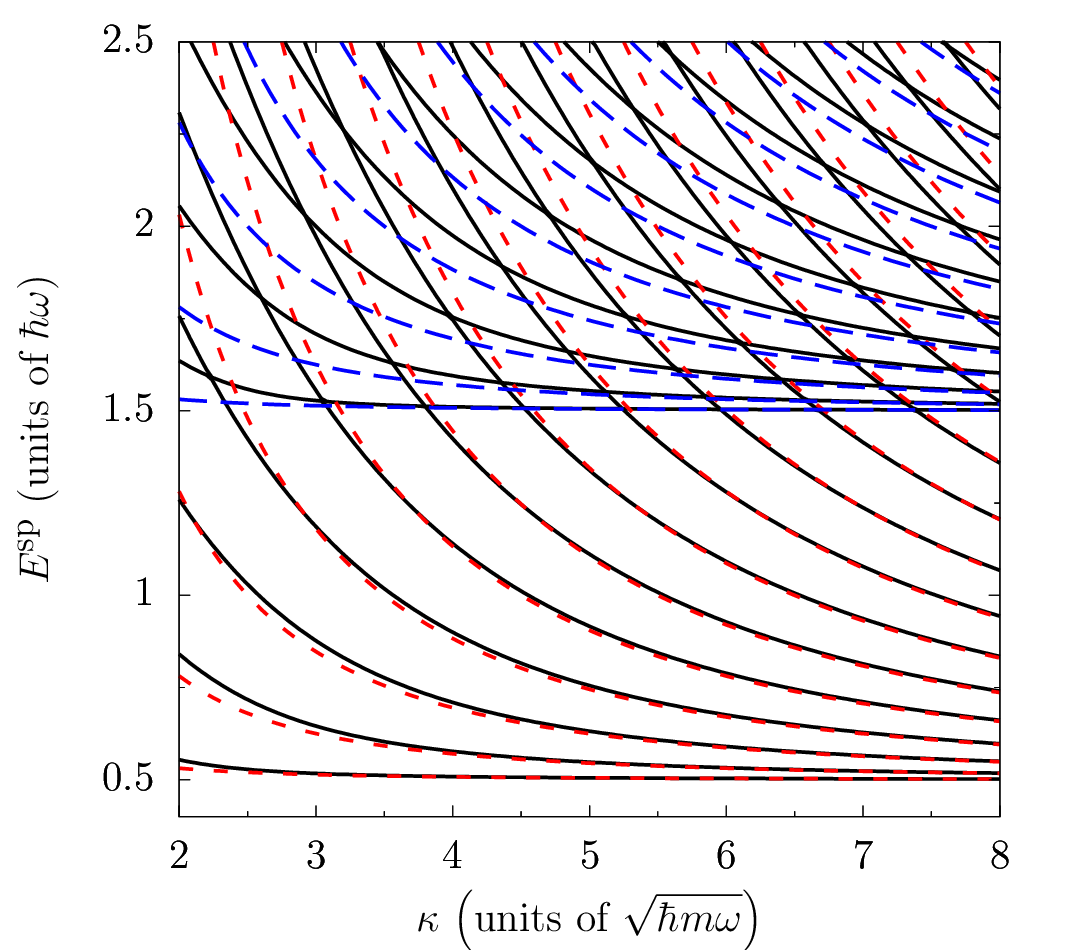}
\caption{\textcolor{black}{Solid black lines: Lowest eigenenergies of the single-particle Hamiltonian in Eq.~(\ref{hamaadxy}) computed by diagonalizing. Short-dashed red lines: Approximate energy levels computed with Eq.~(\ref{Eq.energiesapprox}) and $n_r=0$. Long-dashed blue lines: Approximate energy levels computed with Eq.~(\ref{Eq.energiesapprox}) and $n_r=1$. Notice that each energy level is doubly degenerate and within each kind of lines the energy increases by increasing $j^2_z$.}}
\label{Fig:3}
\end{figure}

The approximate expression, Eq.~(\ref{Eq.energiesapprox}), works 
very well for $\kappa\gg 1$, as seen in Fig.~\ref{Fig:3}. For 
a given value of $\kappa$, the lowest eigenenergies are 
well-described and, as expected, the larger is the value of 
$\kappa$ the better is the approximation for a larger number 
of energy levels.
\subsection{Energy contributions}
\label{Sec.IIC}
As seen above, with increasing $\kappa$ the system goes from a 
harmonic oscillator behavior to a spin-orbit dominated one. The spectral 
properties are very different in both limits and feature a 
particularly involved structure in the intermediate region. 
To better understand the spin-orbit effects, 
we consider now the different energy contributions to the total 
energy of the different eigenstates as we vary the value of $\kappa$. 

\begin{figure}[t!]
\centering
\includegraphics[width=\columnwidth]{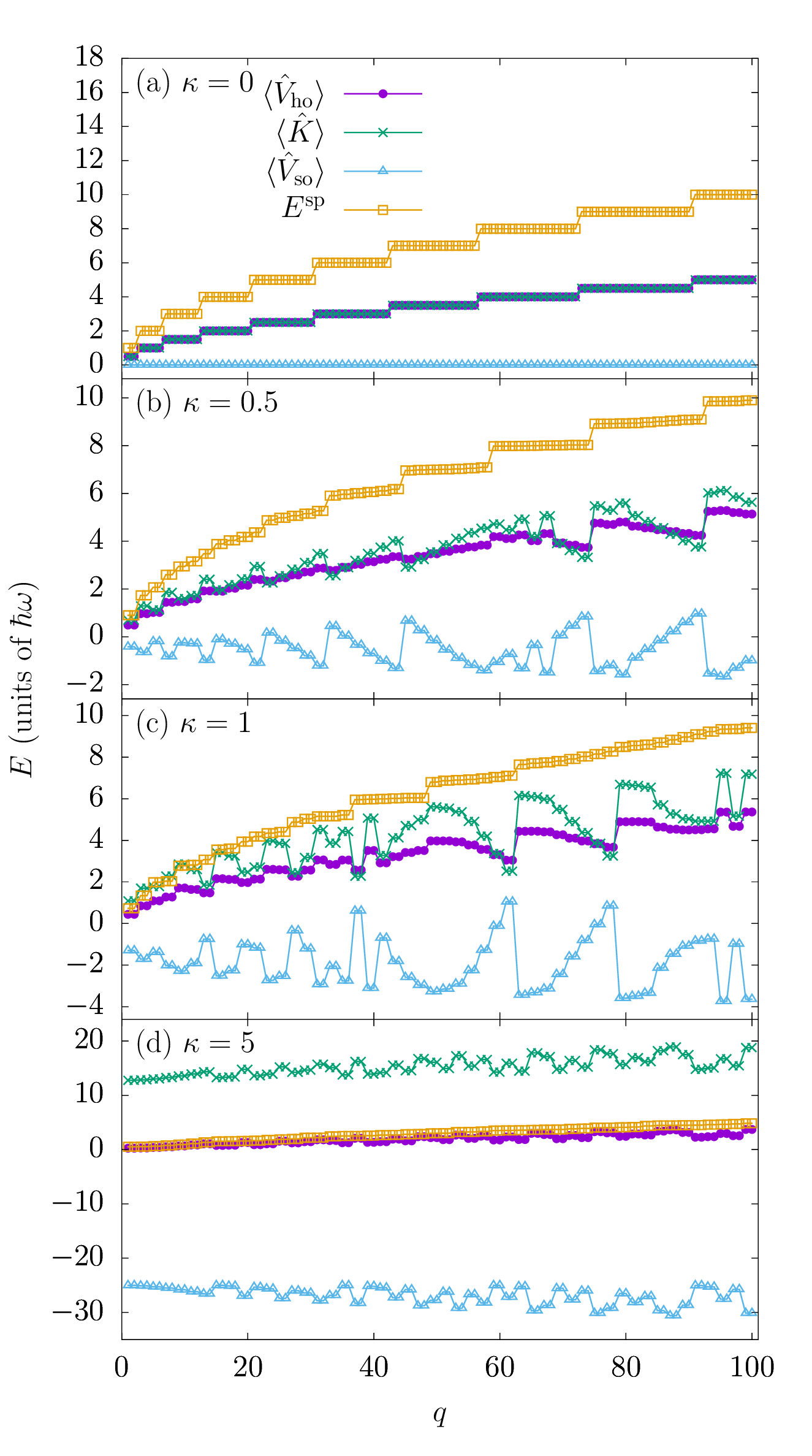}
\caption{\textcolor{black}{Energy contributions to the eigenenergies, $E^{\rm sp}=\langle \hat{K}\rangle+\langle \hat{V}_{\rm so}\rangle+\langle \hat{V}_{\rm ho}\rangle+\frac{\kappa^2}{2}$, for the first $100$ eigenstates of the Hamiltonian in Eq.~(\ref{hamaadxy}), labeled with $q=1,\,...\,,100$. The spin-orbit \textcolor{black}{coupling strength}, $\kappa$, increases going from panel (a) to panel (d). Notice that each panel of this figure corresponds to a vertical cut in Fig.~\ref{Fig:2}. In panel (a), $\langle \hat{K}\rangle$ and $\langle \hat{V}_{\rm ho}\rangle$ coincide.}}
\label{Fig:4}
\end{figure}

In Fig.~\ref{Fig:4}, we show, for the first eigenstates of the single-particle system, how the total energy is distributed between the different energy contributions. As can be seen, the degeneracy due to the time-reversal symmetry of the system, that makes all eigenstates two-fold degenerate, is also reflected in the energy contributions. Each pair of degenerate states has also the same kinetic, harmonic potential, and spin-orbit coupling energies.

In the $\kappa=0$ limit, the eigenstates 
obey the equipartition relation valid for the harmonic oscillator, $\langle K \rangle=\langle V_{\rm ho} \rangle$ [see Fig.~\ref{Fig:4} panel (a)]. For a sufficiently small value of the spin-orbit \textcolor{black}{coupling strength}, those two contributions are not equal but of the same order of magnitude [see 
panels (b) and (c) of Fig.~\ref{Fig:4} for the cases $\kappa=0.5$ and $\kappa=1$, respectively]. Further increasing the value of $\kappa$, the 
situation changes, and the largest contributions, in absolute value, to 
the total energy are clearly the spin-orbit and kinetic parts [see Fig.~\ref{Fig:4} panel (d)]. There are, however, large cancellations 
between these two contributions which result in a total energy 
comparable to the harmonic oscillator part. Further insights into this 
energy decomposition and a nontrivial test to our numerical method is 
provided by the virial theorem (see Appendix~\ref{ap2}), 
\beq
\label{eqvirial}
2\bra{\psi_E^{\rm sp}}\hat{V}_{\rm ho}\ket{\psi_E^{\rm sp}}-2\bra{\psi_E^{\rm sp}}\hat{K}\ket{\psi_E^{\rm sp}}-\bra{\psi_E^{\rm sp}}\hat{V}_{\rm so}\ket{\psi_E^{\rm sp}}=0 \,.
\eeq
For all the states considered, we have checked that the virial theorem energy relation is fulfilled, i.e., the left part of Eq.~(\ref{eqvirial}) represents less than $1\%$ of $E^{\rm sp}$. Actually, the cancellation needed comes 
from $\langle K\rangle$ and $\langle V_{\rm ho}\rangle$ for $\kappa=0$ and from 
$\langle K\rangle$ and $\langle V_{\rm so}\rangle$ in the large 
$\kappa$ domain.

\textcolor{black}{In the presence of spin-orbit coupling, in panels (b)-(d) in Fig.~\ref{Fig:4}, we observe a negative correlation between the spin-orbit coupling term and the kinetic energy. The relation between these two contributions is the following:
\beq
\label{eqkinevso}
\bra{\psi_E^{\rm sp}}\hat{K}\ket{\psi_E^{\rm sp}}=-\frac{3}{4}\bra{\psi_E^{\rm sp}}\hat{V}_{\rm so}\ket{\psi_E^{\rm sp}}+\frac{2E^{\rm sp}-\kappa^2}{4},
\eeq
and it arises from the virial theorem, Eq.~(\ref{eqvirial}), and from writing the energy as:
\beqa
E^{\rm sp}&=&\bra{\psi_E^{\rm sp}}\hat{V}_{\rm ho}\ket{\psi_E^{\rm sp}}+\bra{\psi_E^{\rm sp}}\hat{K}\ket{\psi_E^{\rm sp}}+\bra{\psi_E^{\rm sp}}\hat{V}_{\rm so}\ket{\psi_E^{\rm sp}}
\nonumber \\ 
&+&\kappa^2/2.
\eeqa
}
\subsection{Expectation value of the spin-orbit potential}
\label{Sec.IID}

The term that commonly appears in atomic and nuclear physics as spin-orbit coupling is proportional to $\hat{L}_z \hat{\sigma}_z$. The main difference between that kind of term and the Rasbha spin-orbit is that in one case the spin is coupled to the angular momentum and in the other to the linear momentum. However, we can relate the expectation values of both types of spin-orbit coupling terms,
\beq
\label{eq:virang3}
\bra{\psi_E^{\rm sp}}\hat{V}_{\rm so}\ket{\psi_E^{\rm sp}}=-2\kappa^2\left ( 1+\bra{\psi_E^{\rm sp}}\hat{L}_z\hat{\sigma}_z\ket{\psi_E^{\rm sp}}\right).
\eeq
The eigenstates of the single-particle system obtained by exact diagonalization, whose energies are shown in Fig.~\ref{Fig:4}, fulfill the previous relation, within a numerical error of less than a $1\%$ in the difference between both sides of Eq.~(\ref{eq:virang3}).

The relation between the expectation values of the two kinds of spin-orbit terms is not a particularity of the pure Rashba case, it also works in a more general case, i.e. a mixture of Rashba and Dresselhaus spin-orbit couplings. Moreover, this property does not depend on the external trapping potential. The derivation of the relation in Eq.~(\ref{eq:virang3}) is written in Appendix~\ref{ap2}, where we also generalize it and demonstrate its independence of the external trap.
\section{The two-boson system}
\label{Sec.III}
In this section, we turn to the interacting few-body case. We first 
present our formalism which is developed for the general case of 
$N$ interacting bosons. Afterwards we specialize for the two-boson case. 

Let us thus start with a system of $N$ interacting identical bosons trapped by an isotropic 
harmonic potential with Rashba spin-orbit coupling. The $N$-boson 
Hamiltonian reads
\beq
\label{twobosonHamiltonian}
\hat{H}=\hat{H}_0+\hat{H}_{\rm int}+\frac{N \kappa^2}{2}.
\eeq
The first part contains the total harmonic potential energy, 
$\hat{V}_{\rm ho}^T$, kinetic energy, $\hat{K}^T$, and 
spin-orbit energy, $\hat{V}_{\rm so}^T$,
\beq
\hat{H}_0=\hat{V}_{\rm ho}^T+\hat{K}^T+\hat{V}_{\rm so}^T,
\eeq
with 
$\hat{V}_{\rm ho}^T=(1/2) \sum_{i=1}^N \hat{\bm{x}}^{2}_i$, 
$\hat{K}^T=(1/2)\sum_{i=1}^N \hat{\bm{p}}^{\,2}_i$, and 
$\hat{V}_{\rm so}^T=\kappa \sum_{i=1}^N \left ( \hat{\sigma}_{x_i} \hat{p}_{x_i} +\hat{\sigma}_{y_i} \hat{p}_{y_i} \right )$.

\textcolor{black}{We model the atom-atom interaction with a Gaussian potential~\cite{Mujal},
\beq
\label{eqpotential}
V(|\vec{x}_i-\vec{x}_j|)= \frac{g}{\pi s^2}e^{-\frac{|\vec{x}_i-\vec{x}_j|^2}{s^2}}\,,
\eeq
characterized by a finite range $s$ independent of the spin state and an interaction 
strength, $g$, which can be dependent on the spin~\cite{Yin}. The two interaction parameters are related to the two-dimensional scattering length, $a_{\rm {2D}}$, in Ref.~\cite{Doganov} by comparing the Gaussian potential to zero-range results. The approximate analytical expression that is obtained reads
\beq
a_{\rm {2D}}\approx \sqrt{2}\,s\, e^{-\frac{\gamma}{2}-\frac{2\pi}{g}},
\eeq
where $g$ is written in units of $\hbar^2/m$ and $\gamma$ is the Euler-Mascheroni constant. Beyond that approximation, in Ref.~\cite{Jeszenszki}, the authors combine an analytical treatment with numerical calculations to relate the interaction parameters and $a_{\rm {2D}}$}.

The interaction part is divided in three contributions, 
\beq
\label{eqinteraction}
\hat{H}_{\rm int}=\hat{H}_{\uparrow \uparrow}+\hat{H}_{\downarrow \downarrow}+\hat{H}_{\uparrow \downarrow},
\eeq
where,
\beqa
\hat{H}_{\uparrow \uparrow}&=&\sum_{i<j}^N\frac{g_{\uparrow \uparrow}}{\pi s^2}e^{-\frac{(\hat{\bm{x}}_i-\hat{\bm{x}}_j)^2}{s^2}}\ket{\uparrow}_i\ket{\uparrow}_j\bra{\uparrow}_i\bra{\uparrow}_j, \nonumber\\
\hat{H}_{\downarrow \downarrow}&=&\sum_{i<j}^N\frac{g_{\downarrow \downarrow}}{\pi s^2}e^{-\frac{(\hat{\bm{x}}_i-\hat{\bm{x}}_j)^2}{s^2}}\ket{\downarrow}_i\ket{\downarrow}_j\bra{\downarrow}_i\bra{\downarrow}_j,
\nonumber\\
\hat{H}_{\uparrow \downarrow}&=&\sum_{i<j}^N\frac{g_{\uparrow \downarrow}}{\pi s^2}e^{-\frac{(\hat{\bm{x}}_i-\hat{\bm{x}}_j)^2}{s^2}}\nonumber
\\
&\times&\left(\ket{\uparrow}_i\ket{\downarrow}_j\bra{\uparrow}_i\bra{\downarrow}_j+\ket{\downarrow}_i\ket{\uparrow}_j\bra{\downarrow}_i\bra{\uparrow}_j\right).
\eeqa
For simplicity, we have introduced the following notation for the spin variable: $\ket{\uparrow}\equiv \ket{m_s=1}$, and $\ket{\downarrow}\equiv \ket{m_s=-1}$.
\subsection{Second-quantized two-boson Hamiltonian}
\label{Sec.IIIA}
Despite the fact that our approach is in principle valid for a few number
of bosons, we concentrate from now on in the two-boson case. The two-boson 
system provides a nontrivial example where the interplay of interactions 
and spin-orbit coupling can be studied in detail. 

In our approach we solve numerically the time-independent 
Schr\"odinger equation for the two-boson Hamiltonian truncating the 
Hilbert space. We first consider that the particles can populate 
the first $M$ eigenstates of the harmonic trap, including the 
spin degree of freedom. In this case, we introduce the creation and annihilation operators, $\hat{a}^{\dagger}_i$ and $\hat{a}_i$, that 
create or annihilate bosons in the single-particle state 
$i=1,\,...\,,M$, respectively. They fulfill the 
commutation relations $[\hat{a}_i,\hat{a}^{\dagger}_j]=\delta_{i,j}$ 
and $[\hat{a}_i,\hat{a}_j]=[\hat{a}^{\dagger}_i,\hat{a}^{\dagger}_j]=0$. 
The index $i$ labels the trio of quantum numbers $n_x$, $n_y$ and $m_s$, 
and increases with increasing the energy of the harmonic 
oscillator eigenstate $i$, $\epsilon_{i,i}=n_x(i)+n_y(i)+1$.

The second-quantized version of the single-particle part of Eq.~(\ref{twobosonHamiltonian}) is
\beq
\hat{H}_0=\sum_{i,j=1}^M \hat{a}^{\dagger}_i\hat{a}_j \, \epsilon_{i,j},
\eeq
where the explicit form of $\epsilon_{i,j}$ is given in Eq.~(\ref{ham0matrixelements}). The interaction term is written as:
\beqa
\hat{H}_{\rm int}&=&\frac{1}{2}\sum_{i,j,k,l=1}^M\hat{a}^{\dagger}_i \hat{a}^{\dagger}_j \hat{a}_k \hat{a}_l \, V_{i,j,k,l}
\nonumber\\
&\times & \Bigl\{g_{\uparrow \uparrow} \delta_{m_s(i),1}\delta_{m_s(j),1}\delta_{m_s(k),1}\delta_{m_s(l),1}
\nonumber \\
&+&g_{\downarrow \downarrow}\delta_{m_s(i),-1}\delta_{m_s(j),-1}\delta_{m_s(k),-1}\delta_{m_s(l),-1}
\nonumber\\
&+& g_{\uparrow \downarrow} \Bigl(\delta_{m_s(i),1}\delta_{m_s(j),-1}\delta_{m_s(k),1}\delta_{m_s(l),-1}
\nonumber \\
&+& \delta_{m_s(i),-1}\delta_{m_s(j),1}\delta_{m_s(k),-1}\delta_{m_s(l),1} \Bigr) \Bigr\}\,,
\eeqa
where $V_{i,j,k,l}$ are computed analytically from the 
expressions given in Appendix C of Ref.~\cite{Mujal}, being 
aware that in the present article the indices $i$, $j$, $k$, 
and $l$ label the single-particle states in a different way 
and that the integrals depend on the quantum numbers $n_x$ 
and $n_y$ corresponding to the previous indices.

The Fock states are built creating particles into the vacuum 
state, $\ket{\rm vac}\equiv \ket{0,\,...\,,0}$, as follows:
\beq
\label{Fockbasis}
\ket{n_1,\, ... \, ,n_M}=\frac{(\hat{a}^{\dagger}_1)^{n_1} \dots 
(\hat{a}^{\dagger}_M)^{n_M}}{\sqrt{n_1 !\, ... \, n_M !}} \ket{\rm vac}.
\eeq

In the present work, we study the two-boson case, i.e., 
$
N=\sum_{i=1}^M n_i=2 \,.
\label{eqN2sumq}
$
The basis that we use is the one formed by all the two-boson Fock states with
\beq
\label{truncation}
\sum_{i=1}^M n_i \,\epsilon_{i,i}\leqslant E^{\rm max}=N_E+2,
\eeq
where $N_E$ is a non-negative integer number.
We truncate the Hilbert-space using this energy criterion~\cite{Plodzien}. 
In that case, the Hilbert space dimension considered is given by:
\beq
\label{dimtwoboson}
D(N_E)=\sum_{k=0}^{N_E}\left(3d_{N_E}^b+d_{N_E}^f\right),
\eeq
where $d_{N_E}^b$ and $d_{N_E}^f$ are the number of spatially symmetric and antisymmetric degenerate two-particle states in a two-dimensional 
harmonic trap, given in Eqs. (21) and (22) of Ref.~\cite{Mujal}, respectively, and the factors $3$ and $1$ account for the triplet 
and singlet states of the spin part. The number of modes required 
to accomplish the energy truncation criterion in 
Eq.~(\ref{truncation}) is directly related to $N_E$,
\beq
\label{modesandtrunctwoboson}
M=(N_E+1)(N_E+2) \,.
\eeq

The low-energy eigenstates and eigenenergies of the two-boson 
Hamiltonian matrix are computed numerically using the 
\textit{ARPACK} library. In the following section, we use a Hilbert space of dimension $D=17765$ corresponding to $M=420$ single-particle basis states. In Sec.~\ref{Sec.IV}, we need a larger Hilbert space, with $M=812$ and $D=63035$ [see Eqs.~(\ref{dimtwoboson}) and~(\ref{modesandtrunctwoboson})].
\subsection{Ground-state energy and degeneracy}
In this section, we compute 
the  ground-state energy, concentrating in understanding the way 
the interaction lifts the degeneracy of the ground-state manifold. To 
this aim, we compare our direct diagonalization results with approximate expressions for the energy of the \textcolor{black}{ground-state} manifold. In all cases discussed below, we set the 
spin-orbit coupling to a non-zero but small value, $\kappa=0.3$. Larger 
values of $\kappa$ are discussed \textcolor{black}{in Sec.~\ref{Sec.IV}.}
\begin{figure}
\centering
\includegraphics[width=0.9\columnwidth]{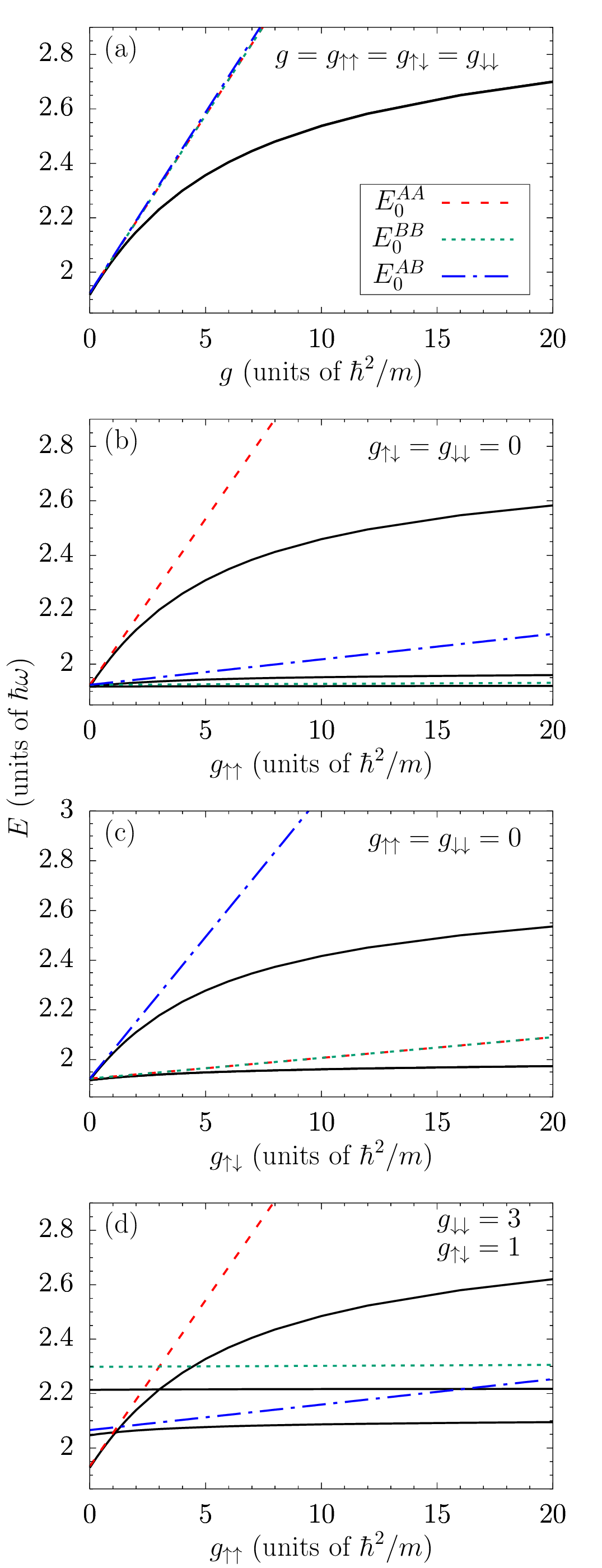}
\caption{\textcolor{black}{(a)-(d) The first three energy levels of the two-boson system depending on the interaction strengths obtained by direct diagonalization (solid black lines).
The approximate perturbative calculations are also plotted [see main text for 
details]. Notice that, in panel (a), the three solid black lines overlap and the dashed and dotted lines, too. Also in the bottom part of panel (c) two solid black lines overlap and two dashed lines, too. We have used a range $s=0.5$ and the spin-orbit \textcolor{black}{coupling strength} $\kappa=0.3$.}}
\label{Fig:5new}
\end{figure}

In absence of interactions, the ground state is three-fold degenerated. 
We obtain approximate 
analytic expressions for the energies of the three states using the six-mode truncation presented \textcolor{black}{in Sec.~\ref{Sec.IIA}.} The energies of the 
three states are denoted, $E_0^{AA}$, $E_0^{AB}$ and $E_0^{BB}$. 
Their explicit expressions are provided in Appendix~\ref{ap1tb}. 

The simplest case we consider is when 
$g_{\uparrow \uparrow}=g_{\downarrow \downarrow}=
g_{\uparrow \downarrow}=g$. In this case, the three orthogonal states 
that define the ground-state subspace remain quasidegenerate [see 
Fig.~\ref{Fig:5new} panel (a)]. As we consider a small finite range, $s=0.5$, the $AB$
state, approximated by Eq.~(\ref{eqstateAB}) at $g\approx 0$, has a 
slightly different energy within our approximation, and would be 
truly degenerate with the other two in the limit of $s\rightarrow 0$.
The three-fold degeneracy of the \textcolor{black}{ground-state} manifold is lifted whenever the interaction strengths are not equal. For instance, fixing $g_{\downarrow \downarrow}=g_{\uparrow \downarrow}=0$, and increasing $g_{\uparrow \uparrow}$ we completely break the degeneracy, since the spin-orbit part of the Hamiltonian induces a nonzero, but different, spin-up spin-up component in all three orthogonal two-boson states. Our perturbative calculations are used to identify which energy level corresponds to each kind of state, as we show in Fig.~\ref{Fig:5new} panel (b). For the case of the state of kind $AA$, the one with a larger spin-up spin-up component, we observe that the prediction of perturbation theory fails for $g_{\uparrow \uparrow}>1$. In contrast, for the state of kind $BB$, with a small spin-up spin-up component, its energy is well-approximated perturbatively up to $g_{\uparrow \uparrow}=20$.

The ground state remains degenerate, although only two-fold, if we set to 
zero the intraspin interactions, $g_{\downarrow \downarrow}
=g_{\uparrow \uparrow}=0$, and vary the inter-spin one, 
$g_{\uparrow \downarrow}$. Since the effect on the states of kind $AA$ 
and $BB$ is the same, they remain degenerate and define the 
ground-state subspace [see Fig.~\ref{Fig:5new} panel (c)]. However, the state $AB$ is 
very sensitive to changes in $g_{\uparrow \downarrow}$, compared to the 
two previous ones, and its energy increases more rapidly.

The last case we consider is fixing at finite values two of the 
interaction strengths, e.g. $g_{\downarrow \downarrow}$ 
and $g_{\uparrow \downarrow}$, and varying the other one, 
$g_{\uparrow \uparrow}$ [see Fig.~\ref{Fig:5new} panel (d)]. In this 
case, we find crossings between the 
energy levels. The perturbative calculations are useful to predict the 
value of $g_{\uparrow \uparrow}$ where the crossing occurs, by 
equating Eqs.~(\ref{eqpertaa}),~(\ref{eqpertbb}) 
and~(\ref{eqpertab}), properly, once $g_{\downarrow \downarrow}$ 
and $g_{\uparrow \downarrow}$ are fixed. In particular, in 
Fig.~\ref{Fig:5new} panel (d), we see that it happens when 
$g_{\uparrow \uparrow}=g_{\downarrow \downarrow}$, and also when 
$g_{\uparrow \uparrow}=g_{\uparrow \downarrow}$.

Finally, we observe that when we further increase the interaction 
strength, regardless of the spin components, the energy levels tend to saturate. This behavior is not captured by the 
perturbative expressions discussed. This is an indicator that the 
system becomes correlated in the proper way in order to reduce the 
total energy by avoiding the atom-atom interaction. This kind of 
behavior was found previously in a harmonically 
trapped system of interacting bosons in two 
dimensions~\cite{Mujal,Mujal2}.
\section{Interaction induced crossover in the $g_{\uparrow \uparrow}
=g_{\downarrow \downarrow}=g_{\uparrow \downarrow}$ case}
\label{Sec.IV}
\begin{figure*}[t!]
\centering
\includegraphics[width=1.6\columnwidth]{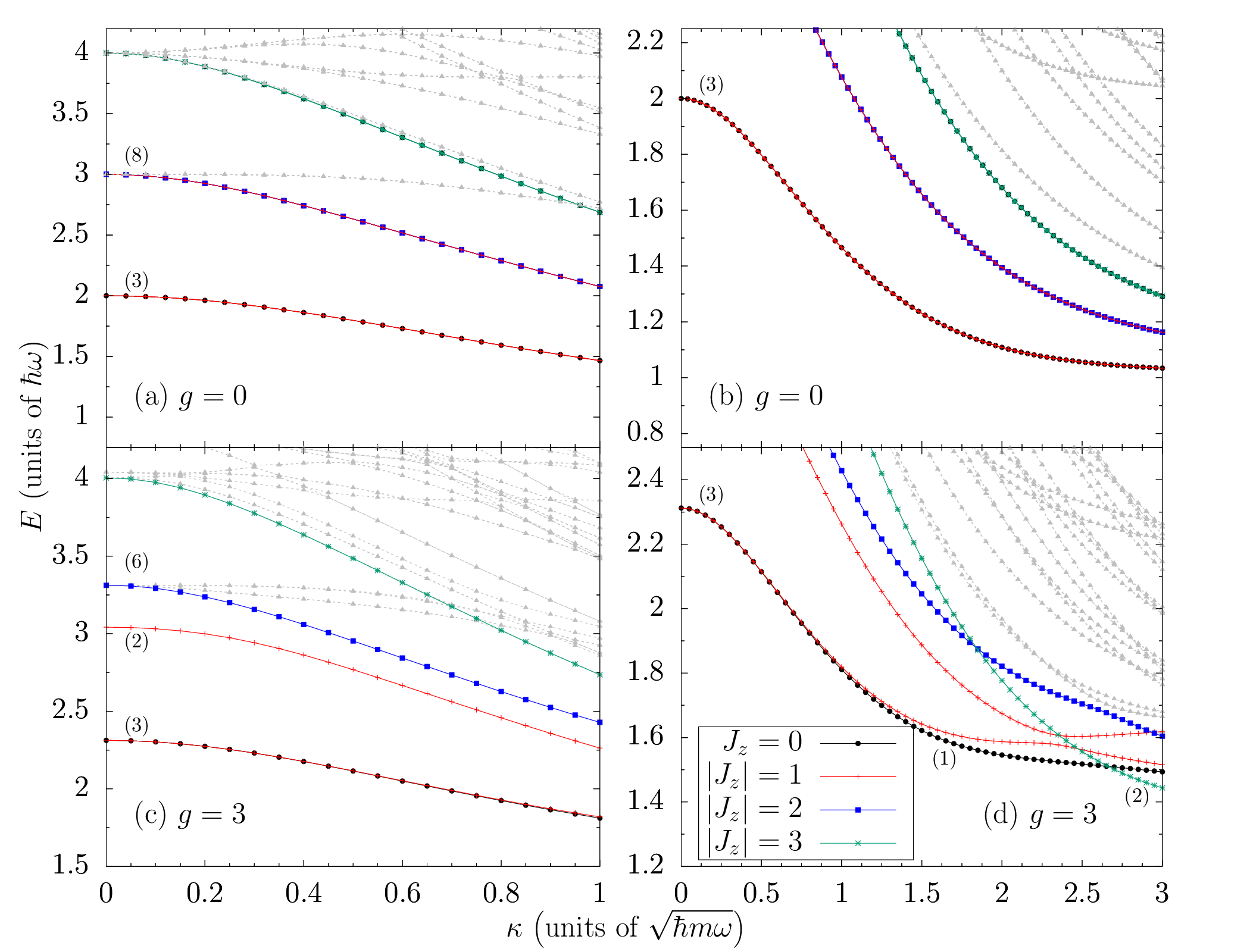}
\caption{\textcolor{black}{Low-energy spectrum of the two-boson system depending on $\kappa$ with $g=0$ in panels (a) and (b), and $g=3$ in panels (c) and (d). Different colors and symbols are used to distinguish energies corresponding to states with different $J_z$. The quantum numbers of the states depicted with grey are not identified in the figure. The energies were computed by diagonalizing, using $M=812$ single-particle basis states that corresponds to a Hilbert-space dimension $D=63035$ [see Eqs.~(\ref{dimtwoboson}) and (\ref{modesandtrunctwoboson})]. We have used a range $s=0.5$. In parenthesis, we provide the degeneracy of the first levels.}}
\label{Fig:6new}
\end{figure*}
Now let us broaden our scope and study not only the ground-state 
manifold but also the lower part of the energy spectrum. The goal
is to discuss the combined effects of the spin-orbit term and the
atom-atom interaction. For simplicity, we consider the case 
$g=g_{\uparrow \uparrow}=g_{\downarrow \downarrow}=g_{\uparrow \downarrow}$, 
with $g\geqslant 0$.

\textcolor{black}{In this situation, the third component of the total angular momentum of the two-particle system,
\beq
\hat{J}_z^T=\hat{J}_z^{(1)}+\hat{J}_z^{(2)},
\eeq
commutes with the Hamiltonian in Eq.~(\ref{twobosonHamiltonian}), so the corresponding quantum number, $J_z$, is a good quantum number to label the eigenstates of $\hat{H}$. Moreover, due to the time-reversal symmetry of the Hamiltonian, the states with $J_z\neq 0$ are at least two-fold degenerate, i.e., with $\pm J_z$. The value of $J_z$ is obtained from the single-particle spectrum in the noninteracting two-particle case and by the numerical diagonalization of $\hat{J}_z^T$ in the degenerate subspaces corresponding to each eigenenergy.}

The interaction has three main effects, as seen in Fig.~\ref{Fig:6new}, where we compare the low energy spectrum for 
$g=0$, panels (a) and (b), with the corresponding ones for $g=3$, panel (c) and (d).
In Fig.~\ref{Fig:6new} panels (a) and (c), we vary $\kappa\in[0,1]$, while in 
panels (b) and (d) we consider a wider region, i.e., $\kappa\in[0,3]$.
Due to the repulsive character of the interaction, the energies are shifted to higher
values, see for instance 
the case of the three-fold degenerate ground-state energy level. A second effect, is the breaking of degeneracies.
For instance, already at $\kappa =0$, the first excited state, with degeneracy 8, breaks in two levels with 
degeneracy 2 for the lowest level and 6 for the highest one. These degeneracies are further broken when 
increasing $\kappa$ [see Fig. ~\ref{Fig:6new} panels (a) and (c)]. Finally, 
the breaking of degeneracies is accompanied by the presence of more 
energy-level crossings, \textcolor{black}{which are avoided between energy levels with the same value of $J_z$ and direct otherwise}.

As seen in Fig.~\ref{Fig:6new} panel (d), we find a crossing at the ground-state level which appears at $\kappa \approx 2.65$ for $g=3$. \textcolor{black}{The numerical calculation of $J_z$ indicates a direct crossing from a nondegenerate ground state with $J_z=0$ to a two-fold degenerate ground-state subspace with two states that can be labeled with $J_z=-3$ and $J_z=3$. Regarding the first-excited energy level, which is two-fold degenerate with $J_z=-1$ and $J_z=1$, it presents an avoided crossing at $\kappa \approx 2.25$.}

\begin{figure}[t!]
\centering
\includegraphics[width=0.78\columnwidth]{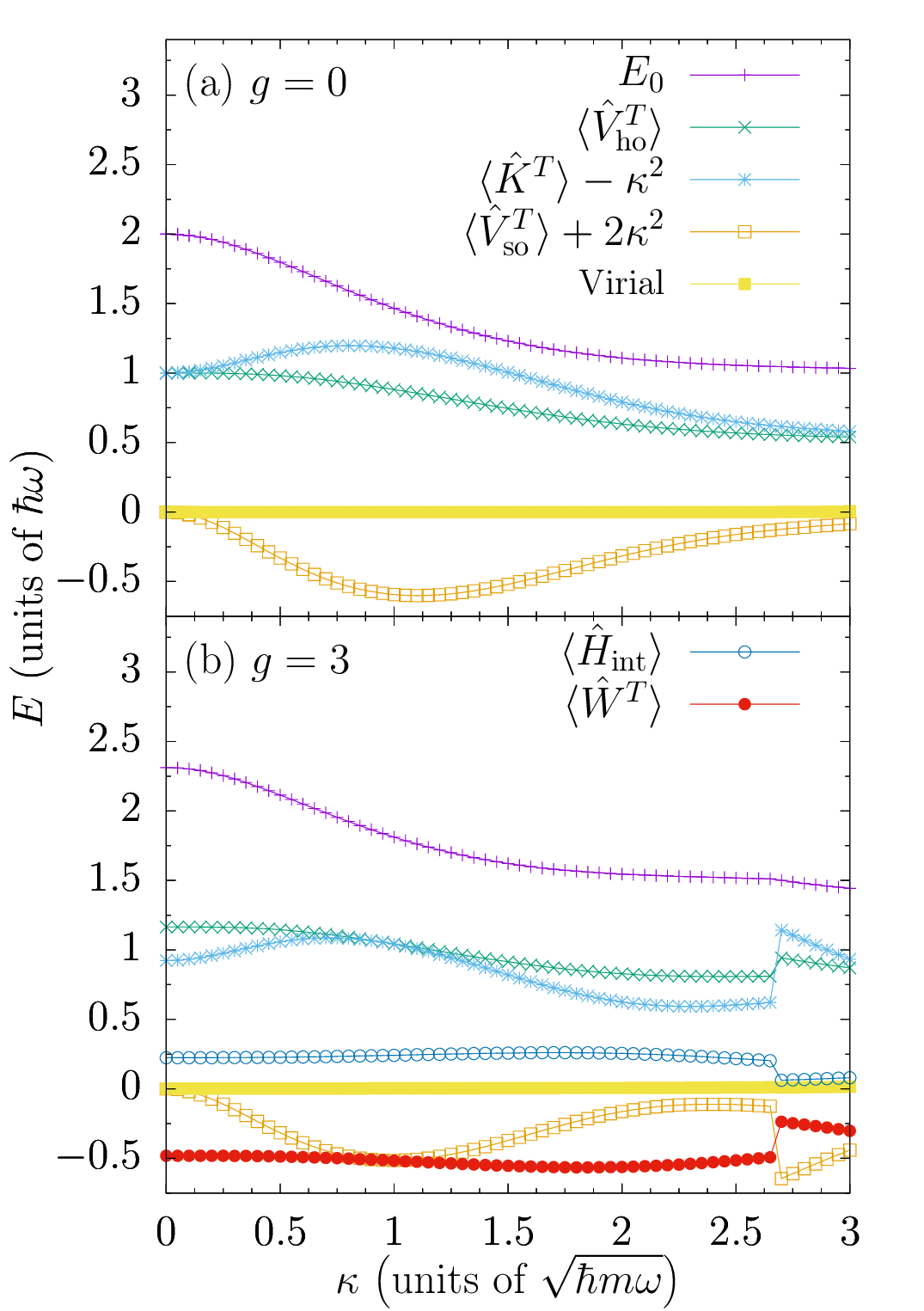}
\caption{\textcolor{black}{The different energy contributions to the two-boson 
ground-state energy and the terms involved in the virial theorem are 
depicted depending on the spin-orbit coupling parameter $\kappa$. In panel (a) $g=0$ and in panel (b) $g=3$. 
${\rm Virial}=2\langle \hat{V}_{\rm ho}^T\rangle
-2\langle \hat{K}^T\rangle-\langle \hat{V}_{\rm so}^T\rangle+\langle \hat{W}^T\rangle$.}}
\label{Fig:15}
\end{figure}

As we will discuss later in Fig.~\ref{Fig:15}, we find a discontinuity in the different energy contributions \textcolor{black}{to the ground-state energy} as expected for a direct crossing.
In the following paragraphs, we concentrate in characterizing this level crossing 
which corresponds to a change in structure of the ground state 
induced by the spin-orbit term in the presence of interactions. 

Starting from $\kappa=0$ and $g=0$, panel (b) of Fig.~\ref{Fig:6new}, 
the \textcolor{black}{ground state} is three-fold degenerate. In this case, one could 
use as a basis of that subspace the two-boson states formed by putting 
the two bosons in the \textcolor{black}{ground state} of the two-dimensional harmonic 
trap with parallel spins, both pointing up or both pointing down, 
and with anti-parallel spins.

For $\kappa > 0$ the previous three states are no longer eigenstates, 
since the spin-orbit imposes a different form for the eigenstates at 
the single-particle level, that was discussed in Sec.~\ref{Sec.IIA}. 
However, the ground-state degeneracy remains unchanged with 
increasing $\kappa$ in the noninteracting case. The three states that 
define the ground-state subspace are
\beq
\label{groundAAtwo}
\ket{\Psi_{0,AA}}=\ket{\psi^{\rm sp}_{0,A}}\ket{\psi^{\rm sp}_{0,A}},
\eeq
\beq
\label{groundBBtwo}
\ket{\Psi_{0,BB}}=\ket{\psi^{\rm sp}_{0,B}}\ket{\psi^{\rm sp}_{0,B}},
\eeq
and
\beq
\label{groundABtwo}
\ket{\Psi_{0,AB}}=\frac{1}{\sqrt{2}}\left(\ket{\psi^{\rm sp}_{0,A}}\ket{\psi^{\rm sp}_{0,B}}+\ket{\psi^{\rm sp}_{0,B}}\ket{\psi^{\rm sp}_{0,A}}\right),
\eeq
constructed with the two-degenerate single-particle eigenstates, 
$\ket{\psi^{\rm sp}_{0,A}}$ and $\ket{\psi^{\rm sp}_{0,B}}$, of the Hamiltonian in Eq.~(\ref{eq1ham}). 

In the interacting case the three-fold degenerate 
ground-state subspace splits in two energy levels: the ground state 
becomes nondegenerate\textcolor{black}{, with $J_z=0$}, and the first excitation becomes two-fold 
degenerate\textcolor{black}{, corresponding to one state with $J_z=-1$ and the other with $J_z=1$}. This effect is more notorious for larger 
$\kappa$, for instance for $\kappa=1.5$ in 
Fig.~\ref{Fig:6new} panel (d), where we observe the gap opening. For 
larger $\kappa$ we observe the previously mentioned crossing. 
From $\kappa \approx 2.65$ up to $3$, the ground state becomes 
two-fold degenerate \textcolor{black}{with $J_z=-3$ and $J_z=3$, respectively. The level which crosses at $\kappa\approx 2.65$ corresponds to the evolution with $\kappa$ of an excited level with $E=4$ at $\kappa=0$ that directly crosses multiple levels [see Fig.~\ref{Fig:6new} panels (c) and (d)].} Let us emphasize 
that this transition is a joint effect of the spin-orbit coupling 
and the interaction, since it is only observed when both effects 
are present.

To characterize the crossing in the ground-state energy we have 
computed its energy contributions in the cases of Fig.~\ref{Fig:6new} 
panels (b) and (d). These results are shown in Fig.~\ref{Fig:15}, where we have also tested the fulfillment of the 
virial theorem energy relation (see Appendix~\ref{ap2}).

Before the crossing, the dependence on $\kappa$ of the kinetic, 
the harmonic potential and the spin-orbit energies is 
qualitatively similar to the noninteracting case (see Fig.~\ref{Fig:15}).
In the interacting case, the atoms are farther 
from the center of the trap resulting in a shift in the harmonic 
potential energy between the $g=0$ and $g=3$ cases depicted in 
Fig.~\ref{Fig:15}. The kinetic energy is reduced in the interacting 
case. The interaction energy and the term coming from the interaction 
present in the virial relation, $\langle \hat{W}^T \rangle$, are 
mostly independent of $\kappa$. \textcolor{black}{This fact explains that
the correlations between the spin-orbit coupling energy and the kinetic energy,
given in Eq.~(\ref{eqkinesonboson}), are similar to the noninteracting case.}
At the crossing, except from 
the total energy that remains continuous, all other energy terms 
feature a discontinuity. After the crossing, the ground state has 
a different structure. \textcolor{black}{The harmonic potential and the kinetic energies are 
larger than before. Again, this positive terms are compensated by 
the negative spin-orbit term that is larger in absolute value.}
In particular, the state is less sensitive 
to the presence of the repulsive interaction, since the interaction 
energy is smaller and closer to zero compared to the other energy 
terms.
\begin{figure}[t!]
\centering
\includegraphics[width=0.9\columnwidth]{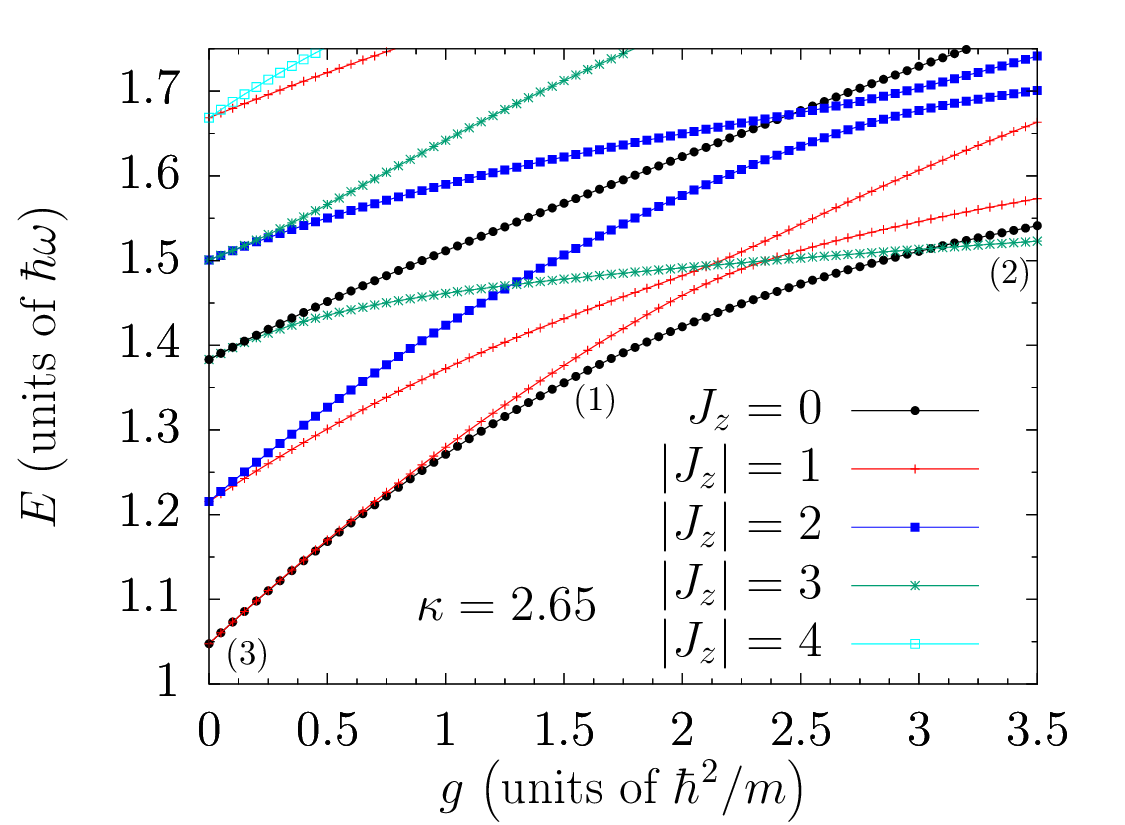}
\caption{\textcolor{black}{Low-energy spectrum for the two-boson system at $\kappa=2.65$ depending on the interaction strength $g$. The energies were computed by diagonalizing, using $M=812$ single-particle basis states that corresponds to a Hilbert-space dimension $D=63035$ [see Eqs.~(\ref{dimtwoboson}) and (\ref{modesandtrunctwoboson})]. We have used a range $s=0.5$.  In parenthesis, we provide the degeneracy of the first levels.}}
\label{Fig:new9}
\end{figure}
\begin{figure}[b!]
\centering
\includegraphics[width=0.9\columnwidth]{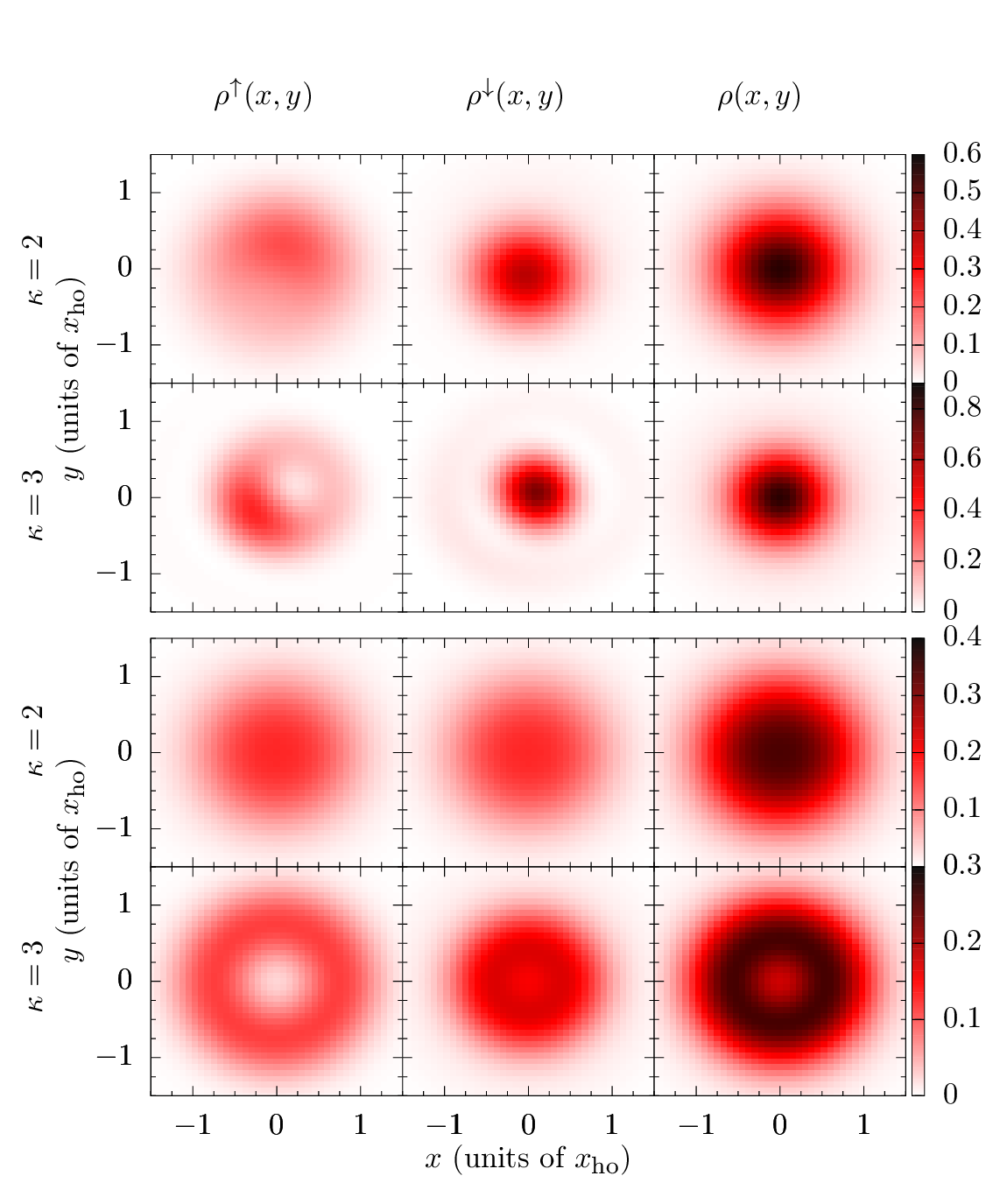}
\caption{\textcolor{black}{Density profiles of each spin-component and the total one for $\kappa=2$ and $\kappa=3$. 
The upper and lower panels correspond to the noninteracting, $g=0$, 
and interacting, $g=3$, cases, respectively.}}
\label{Fig:12}
\end{figure}

\textcolor{black}{This fact is better observed in Fig.~\ref{Fig:new9}, where the
crossing is found varying $g$ with fixing $\kappa$. The energy level that finally becomes
the ground state is one of the levels forming the third-excited manifold at $g=0$. This level directly crosses
with multiple levels when $g$ is increased since its energy is less sensitive to the increase of the interaction strength value. In parallel, there are avoided crossings between the states that have the same quantum number $J_z$, as in the cases of $|J_z|=1$ and $|J_z|=2$, at $g\approx 2$ and $g\approx 3$, respectively.}  

The effects of the crossover become also apparent in the 
density of the cloud (see Appendix~\ref{app:den} for the explicit expressions). To illustrate this phenomenology we compare 
the densities for the $g=0$ and $g=3$ cases, for two values before and
after the level crossing, $\kappa=2$ and $\kappa=3$, respectively. 
For $\kappa=2$ we observe that the total density of the cloud is 
similar in both cases (see Fig.~\ref{Fig:12}). The main difference is that the interacting 
cloud is already larger than the noninteracting one, as expected 
from the repulsive nature of the atom-atom interactions considered. 
The densities of the two spin components are different for $g=0$ and $g=3$. 
In the interacting case, both densities are very similar, while in the 
noninteracting one $\rho^\downarrow$ is much smaller and more peaked at 
the center of the trap. 
An important effect of the crossing is that the cloud becomes larger 
after the level crossing, i.e. going from $\kappa=2$ to 
$\kappa=3$ for $g=3$ (see the total density in Fig.~\ref{Fig:12}). 
This is in contrast with the behavior observed in absence of 
interactions, where the cloud size gets reduced when going from 
$\kappa=2$ to $\kappa=3$, as seen in Fig.~\ref{Fig:12}. This effect is 
observed also for the densities of each component separately. 
Another relevant feature is that, after the crossing, the total 
density has a dip in the center of the trap, while in 
the noninteracting case it has a maximum. 

\section{Summary and Conclusions}
\label{Sec.V}
We have considered one and two bosons trapped in a harmonic potential with the presence of spin-orbit coupling.
For the single-particle case, the diagonalization of the Hamiltonian matrix has allowed us to study the properties of the low-energy eigenstates of the system, going from the weak spin-orbit coupling regime to the strong one. We have computed the expectation values of each energy term in the Hamiltonian for the eigenstates, separately, and have derived and tested the virial energy relation between them. In particular, we have found a relation between the expectation value of different kind of spin-orbit coupling terms, which is independent of the trapping potential. For the ground state of the single-particle system, we have derived approximate analytical expressions that are able to reproduce the ground-state energy in the weak spin-orbit coupling regime and that, for the interacting two-boson system, are used to obtain perturbative expressions that explain the breaking of the degeneracy of the ground-state subspace when changing the values of the spin-dependent interaction strengths. In all cases, we have found that the ground-state energy tends to saturate with increasing the strength of the interaction, departing from the perturbation-theory prediction. This signals the formation of repulsive correlations in the system. In addition, in the spin-independent interaction case, for the repulsively interacting two-boson system, \textcolor{black}{we have found a direct crossing in the ground state corresponding to states with different values of $J_z$, from $J_z=0$ to $J_z=\pm 3$. Some avoided crossings between excited states with the same $J_z$ are observed} when the spin-orbit coupling parameter is sufficiently large. The change in the ground state has been characterized by computing the energy contributions, that present a discontinuity at the point where there is a \textcolor{black}{direct} energy-level crossing in the ground-state energy. Moreover, this phenomenon has been observed to be apparent in the density profile of the system, which could be experimentally
measured~\cite{Pyzh,McDonald,Subhankar}.
\vspace*{0.5cm}
\begin{acknowledgments}
Useful discussion with Gordon Baym at the early stage of the work are gratefully acknowledged.
We also thank Ferran Mazzanti and Juan S\'anchez-Baena for their comments and for sharing their results with us.
P.M. wants to thank Doerte Blume for giving him the opportunity to visit her group and thank her and Qingze Guan, Jianwen Jie and Jugal Talukdar for their warm hospitality and 
	useful discussions.
We acknowledge financial support from the Spanish Ministerio de Economia y Competitividad Grant No FIS2017-87534-P, from Generalitat de Catalunya Grant No. 2017SGR533 and from the European Union Regional Development Fund within the ERDF Operational Program of Catalunya (project QUASICAT/QuantumCat).
P.M. was supported by a FI grant from Generalitat de Catalunya.
\end{acknowledgments}
\appendix
\section{Analytical approximations in the weak spin-orbit coupling regime}

\subsection{Single-particle case}
\label{ap1}

In a first approximation, we consider a Hilbert space of dimension $6$, where the particle can populate the ground state of the harmonic oscillator or one of the two first-excited states of the trap, considering also the two possible spin orientations. Therefore, we consider the basis $\{\ket{n_x,n_y,m_s}\}=\{\ket{0,0,1},\ket{0,0,-1},\ket{1,0,1},\ket{1,0,-1},\ket{0,1,1},\ket{0,1,-1}\}$. In this Hilbert space, we construct the Hamiltonian matrix and diagonalize it analytically with \textit{Mathematica}. In this way, we find approximate expressions for the ground state and its energy depending on the spin-orbit \textcolor{black}{coupling strength}, $\kappa$. The single-particle  ground-state energy is approximately given by, 
\beq
E_{0,d=6}^{\rm sp}= \frac{1}{2}\left(3-\sqrt{4\kappa^2+1}\right)+\frac{\kappa^2}{2} \,. 
\eeq
The ground state is two-fold degenerate, and we label with $A$ and $B$ the orthogonal states,
\beq
\label{eqap1_2}
\ket{\psi^{\rm sp}_{0,A}}_{d=6} = -C_0\ket{0,0,1}
+C_1\left(i\ket{1,0,-1}-\ket{0,1,-1}\right),
\eeq
and
\beq
\label{eqap1_3}
\ket{\psi^{\rm sp}_{0,B}}_{d=6}= C_0\ket{0,0,-1}
+C_1\left(-i\ket{1,0,1}-\ket{0,1,1}\right),
\eeq
where $C_0$ and $C_1$ are given by
\beq
\label{eqap1_4}
C_0(\kappa)=\frac{\kappa \sqrt{4+\frac{1+\sqrt{1+4\kappa^2}}{\kappa^2}}}{\sqrt{2+8\kappa^2}},
\eeq
and
\beq
\label{eqap1_5}
C_1(\kappa)=\frac{1}{\sqrt{4+\frac{1+\sqrt{1+4\kappa^2}}{\kappa^2}}}.
\eeq

Repeating the previous procedure with a Hilbert space of dimension $12$, we obtain more accurate expressions for the  ground-state energy, given by,
\beq
E_{0,d=12}^{\rm sp}=2-\sqrt{2\kappa^2+1}+\frac{\kappa^2}{2}.
\eeq
and also for the coefficients of the two degenerate states
\beq
\label{eqap1_7}
\begin{gathered}
\ket{\psi^{\rm sp}_{0,A}}_{d=12}= -D_0\ket{0,0,1}+D_1\left(i\ket{1,0,-1}-\ket{0,1,-1}\right)
\\
+D_2\left(\ket{0,2,1}+\ket{2,0,1}\right),
\end{gathered}
\eeq
and
\beq
\label{eqap1_8}
\begin{gathered}
\ket{\psi^{\rm sp}_{0,B}}_{d=12}= D_0\ket{0,0,-1}+D_1\left(-i\ket{1,0,1}-\ket{0,1,1}\right)
\\
-D_2\left(\ket{0,2,-1}+\ket{2,0,-1}\right),
\end{gathered}
\eeq
where $D_0$, $D_1$ and $D_2$ are given by
\beq
\label{eqap1_9}
D_0(\kappa)=\sqrt{\frac{\kappa^2+1+\sqrt{2\kappa^2+1}}{4\kappa^2+2}},
\eeq
\beq
\label{eqap1_10}
D_1(\kappa)=\frac{\kappa \left(1+\sqrt{2\kappa^2+1}\right)}{2\sqrt{\left(2\kappa^2+1\right)\left(\kappa^2+1+\sqrt{2\kappa^2+1}\right)}},
\eeq
and
\beq
\label{eqap1_11}
D_2(\kappa)=\frac{\kappa^2}{2\sqrt{\left(2\kappa^2+1\right)\left(\kappa^2+1+\sqrt{2\kappa^2+1}\right)}}.
\eeq

\subsection{Two-boson case}
\label{ap1tb}

Within the first single-particle approximation for small $\kappa$, discussed in Sec.~\ref{Sec.IIA}, we compute the energy of the following two-boson states:
\beq
\label{eqstateAA}
\ket{\Phi_{0,AA}}=\ket{\psi^{\rm sp}_{0,A}}_{d=6}\ket{\psi^{\rm sp}_{0,A}}_{d=6},
\eeq
\beq
\label{eqstateBB}
\ket{\Phi_{0,BB}}=\ket{\psi^{\rm sp}_{0,B}}_{d=6}\ket{\psi^{\rm sp}_{0,B}}_{d=6},
\eeq
and
\beqa
\ket{\Phi_{0,AB}}&=&\frac{1}{\sqrt{2}}\left(
\ket{\psi^{\rm sp}_{0,A}}_{d=6}\ket{\psi^{\rm sp}_{0,B}}_{d=6}\right.
\nonumber\\
&+&\left.\ket{\psi^{\rm sp}_{0,B}}_{d=6}\ket{\psi^{\rm sp}_{0,A}}_{d=6}\right),
\label{eqstateAB}
\eeqa
up to first order in perturbation theory for the interaction 
strength parameters $g_{\uparrow \uparrow}$, $g_{\downarrow \downarrow}$, 
and $g_{\uparrow \downarrow}$. The previous three states 
describe, approximately, the degenerate two-boson ground-state subspace 
in the noninteracting limit. The approximation becomes exact in the 
limit of $\kappa \rightarrow 0$. The first part of the energy for all 
of them is computed multiplying the single-particle energy given 
in Eq.~(\ref{eqtrunc6}) by the number of particles, that is $2$. 
The interaction part arises from computing the expectation values 
$\bra{\Phi_{0,AA}}\hat{H}_{\rm int}\ket{\Phi_{0,AA}}$, $\bra{\Phi_{0,BB}}\hat{H}_{\rm int}\ket{\Phi_{0,BB}}$, and 
$\bra{\Phi_{0,AB}}\hat{H}_{\rm int}\ket{\Phi_{0,AB}}$, since 
$\bra{\Phi_{0,AA}}\hat{H}_{\rm int}\ket{\Phi_{0,BB}}=\bra{\Phi_{0,AA}}\hat{H}_{\rm int}\ket{\Phi_{0,AB}}=\bra{\Phi_{0,BB}}\hat{H}_{\rm int}\ket{\Phi_{0,AB}}=0$. Therefore, the energies are
\beq
\begin{gathered}
\label{eqpertaa}
E^{AA}_0 = 3-\sqrt{4\kappa^2+1}+\kappa^2+\frac{g_{\uparrow \uparrow}C^4_0}{\pi (2+s^2)}
\\
+\frac{g_{\downarrow \downarrow}4 C^4_1 (2+2s^2+s^4)}{\pi (2+s^2)^3}+\frac{g_{\uparrow \downarrow}4C^2_0 C^2_1}{\pi (2+s^2)^2},
\end{gathered}
\eeq
\beq
\begin{gathered}
\label{eqpertbb}
E^{BB}_0 = 3-\sqrt{4\kappa^2+1}+\kappa^2+\frac{g_{\downarrow \downarrow}C^4_0}{\pi (2+s^2)}
\\
+\frac{g_{\uparrow \uparrow}4 C^4_1 (2+2s^2+s^4)}{\pi (2+s^2)^3}+\frac{g_{\uparrow \downarrow}4C^2_0 C^2_1}{\pi (2+s^2)^2},
\end{gathered}
\eeq
and
\beq
\begin{gathered}
\label{eqpertab}
E^{AB}_0 = 3-\sqrt{4\kappa^2+1}+\kappa^2+\frac{\left(g_{\uparrow \uparrow}+g_{\downarrow \downarrow}\right)2C^2_0C^2_1}{\pi (2+s^2)}
\\
+g_{\uparrow \downarrow}\left(\frac{C^4_0}{\pi (2+s^2)}-\frac{4C^2_0 C^2_1}{\pi (2+s^2)^2}+\frac{8C^4_1}{\pi (2+s^2)^3}\right),
\end{gathered}
\eeq
where $C_0$ and $C_1$ depend on $\kappa$ and are given in Eq.~(\ref{eqap1_4}) and Eq.~(\ref{eqap1_5}) of Appendix~\ref{ap1}, respectively.

A particular limit case of interest is the short-range limit, $s\rightarrow 0$. In that case, the previous expressions reduce to
\beq
\begin{gathered}
E^{AA}_{0,s\rightarrow 0}= 3-\sqrt{4\kappa^2+1}+\kappa^2
\\
+\frac{g_{\uparrow \uparrow}C^4_0+g_{\uparrow \downarrow}2C^2_0 C^2_1+g_{\downarrow \downarrow}2C^4_1}{2\pi},
\end{gathered}
\eeq
\beq
\begin{gathered}
E^{BB}_{0,s\rightarrow 0}= 3-\sqrt{4\kappa^2+1}+\kappa^2
\\
+\frac{g_{\downarrow \downarrow}C^4_0+g_{\uparrow \downarrow}2C^2_0 C^2_1+g_{\uparrow \uparrow}2C^4_1}{2\pi},
\end{gathered}
\eeq
and
\beq
\begin{gathered}
E^{AB}_{0,s\rightarrow 0}= 3-\sqrt{4\kappa^2+1}+\kappa^2
\\
+\frac{g_{\uparrow \downarrow}\left(C^4_0+2C^4_1\right)+\left(g_{\uparrow \uparrow}+g_{\downarrow \downarrow}-g_{\uparrow \downarrow}\right)2C^2_0C^2_1}{2\pi}.
\end{gathered}
\eeq
\section{Virial relations}
\label{ap2}
\subsection{Virial theorem energy relation}
\label{ap2_1}
For the eigenstates, $\ket{\Psi_E}$, of the Hamiltonian in Eq.~(\ref{twobosonHamiltonian}), i.e., $\hat{H}\ket{\Psi_E}=E\ket{\Psi_E}$, the virial theorem establishes that
\beq
\label{eq1_ap2}
\begin{gathered}
\bra{\Psi_E}[\hat{H},\hat{\mathcal{O}}^T]\ket{\Psi_E}=
\\
=\bra{\psi_E}\left (\hat{H}\hat{\mathcal{O}}^T-\hat{\mathcal{O}}^T\hat{H}\right)\ket{\psi_E}
\\
= \bra{\psi_E}\left (E\hat{\mathcal{O}}^T-\hat{\mathcal{O}}^T E\right)\ket{\psi_E}=0,
\end{gathered}
\eeq
with $\hat{\mathcal{O}}^T=\sum_{i=1}^N\left(\hat{x}_i\hat{p}_{x_i}+\hat{y_i}\hat{p}_{y_i}\right)$. The explicit computation of the expectation value of the commutator on the left part of the previous equation results in:
\beq
\label{eq2_ap2}
\begin{gathered}
2\bra{\Psi_E}\hat{V}_{\rm ho}^T\ket{\Psi_E}-2\bra{\Psi_E}\hat{K}^T\ket{\Psi_E}-\bra{\Psi_E}\hat{V}_{\rm so}^T\ket{\Psi_E}
\\
+\bra{\Psi_E}\hat{W}^{\uparrow \uparrow}\ket{\Psi_E}+\bra{\Psi_E}\hat{W}^{\uparrow \downarrow}\ket{\Psi_E}+\bra{\Psi_E}\hat{W}^{\downarrow \downarrow}\ket{\Psi_E}=0,
\end{gathered}
\eeq
where the last three terms come from the interaction part of the Hamiltonian~(\ref{eqinteraction}) and the operators involved read:
\beq
\label{eq3_ap2}
\hat{W}^{\uparrow \uparrow}=-\sum_{i<j}^N\frac{2g^{\uparrow \uparrow}}{\pi s^4}(\hat{\bm{x}}_i-\hat{\bm{x}}_j)^2e^{-\frac{(\hat{\bm{x}}_i-\hat{\bm{x}}_j)^2}{s^2}}\ket{\uparrow}_i\ket{\uparrow}_j\bra{\uparrow}_i\bra{\uparrow}_j,
\eeq
\beq
\label{eq4_ap2}
\hat{W}^{\downarrow \downarrow}=-\sum_{i<j}^N\frac{2g^{\downarrow \downarrow}}{\pi s^4}(\hat{\bm{x}}_i-\hat{\bm{x}}_j)^2e^{-\frac{(\hat{\bm{x}}_i-\hat{\bm{x}}_j)^2}{s^2}}\ket{\downarrow}_i\ket{\downarrow}_j\bra{\downarrow}_i\bra{\downarrow}_j,
\eeq
and
\beq
\label{eq5_ap2}
\begin{split}
\hat{W}^{\uparrow \downarrow}=&-\sum_{i<j}^N\frac{2g^{\uparrow \downarrow}}{\pi s^4}(\hat{\bm{x}}_i-\hat{\bm{x}}_j)^2e^{-\frac{(\hat{\bm{x}}_i-\hat{\bm{x}}_j)^2}{s^2}}
\\
\times&\left(\ket{\uparrow}_i\ket{\downarrow}_j\bra{\uparrow}_i\bra{\downarrow}_j+\ket{\downarrow}_i\ket{\uparrow}_j\bra{\downarrow}_i\bra{\uparrow}_j\right).
\end{split}
\eeq
We also define the operator:
\beq
\label{eq6_ap2}
\hat{W}^T \equiv \hat{W}^{\uparrow \uparrow}+\hat{W}^{\downarrow \downarrow}+\hat{W}^{\uparrow \downarrow}.
\eeq

\textcolor{black}{Additional relations between the energy contributions are derived from the virial theorem in Eq.~(\ref{eq2_ap2}) and the fact that the total energy is given by:
\beqa
E&=&\bra{\Psi_E}\hat{V}_{\rm ho}^T\ket{\Psi_E}+\bra{\Psi_E}\hat{K}^T\ket{\Psi_E}+\bra{\Psi_E}\hat{V}_{\rm so}^T\ket{\Psi_E}
\nonumber \\
&+& \bra{\Psi_E}\hat{H}_{\rm {int}}\ket{\Psi_E}+\frac{N\kappa^2}{2}.
\eeqa
For instance, the generalization of Eq.~(\ref{eqkinevso}) in the presence of interactions for the $N$-particle system, that relates the spin-orbit coupling term and the kinetic energy, reads
\beqa
\label{eqkinesonboson}
\bra{\Psi_E}\hat{K}^T\ket{\Psi_E}=-\frac{3}{4}\bra{\Psi_E}\hat{V}_{\rm so}^T\ket{\Psi_E}+\frac{2E-N\kappa^2}{4}
\nonumber \\
-\frac{1}{2}\bra{\Psi_E}\hat{H}_{\rm {int}}\ket{\Psi_E}+\frac{1}{4}\bra{\Psi_E}\hat{W}^T\ket{\Psi_E} .
\eeqa
}

In the noninteracting case, with the relation in Eq.~(\ref{eq2_ap2}) we can write the eigenenergies of the Hamiltonian in Eq.~(\ref{twobosonHamiltonian}) as:
\beq
\label{eq7_ap2}
E=3\bra{\Psi_E}\hat{V}_{\rm ho}^T\ket{\Psi_E}-\bra{\Psi_E}\hat{K}^T\ket{\Psi_E}+\frac{N\kappa^2}{2}.
\eeq

In the single-particle case, the virial theorem energy relation, Eq.~(\ref{eq2_ap2}), reduces to Eq.~(\ref{eqvirial}).
\subsection{Angular momenta and spin-orbit virial relation}
Following the same procedure of previous Sec.~\ref{ap2_1}, we compute the expectation value of the following commutator:
\beq
\label{eq8_ap2}
\bra{\Psi_E}[\hat{H}^{RD},\hat{\mathcal{O}}^T]\ket{\Psi_E}=0,
\eeq
with $\hat{\mathcal{O}}^T=\sum_{i=1}^N\kappa\left(\hat{x}_i\hat{\sigma}_{x_i}+\eta \hat{y_i}\hat{\sigma}_{y_i}\right)$. In this case, we have used the general many-body Hamiltonian, that describes a noninteracting system,
\beq
\label{eq9_ap2}
\hat{H}^{RD}=\hat{V}^T+\hat{K}^T+\hat{V}_{\rm so}^{RD,T},
\eeq
where the external trap is an arbitrary potential of the form
\beq
\label{eq10_ap2}
\hat{V}^T=\sum_{i=1}^N\hat{V}(\hat{x_i},\hat{y_i}),
\eeq
and the spin-orbit term is a mixture of Rashba and Dresselhaus of the form:
\beq
\label{eq11_ap2}
\hat{V}_{\rm so}^{RD,T}=\kappa\sum_{i=1}^N\left ( \hat{\sigma}_{x_i} \hat{p}_{x_i} +\eta \hat{\sigma}_{y_i} \hat{p}_{y_i} \right ).
\eeq

As a result, we find that
\beq
\label{eq12_ap2}
\begin{gathered}
\bra{\Psi_E}\hat{V}^{RD,T}_{\rm so}\ket{\Psi_E}
\\
 =-\kappa^2\Big(N\left(1+\eta^2\right)+2\eta\bra{\Psi_E}\sum_{i=1}^N\hat{L}_{z_i}\hat{\sigma}_{z_i}\ket{\Psi_E}\Big),
\end{gathered}
\eeq
where now, $\ket{\Psi_E}$ are the eigenstates of $\hat{H}^{RD}$.
The independence of the external trapping potential arises from the fact that
\beq
\label{eq13_ap2}
[\hat{V}^T,\hat{\mathcal{O}}^T]=0.
\eeq
In the single-particle case and with a pure Rashba-type spin-orbit coupling the relation of Eq.~(\ref{eq12_ap2}) is equivalent to Eq.~(\ref{eq:virang3}).
\section{Densities}
\label{app:den}
The total density is computed as the expectation value of the operator
\begin{equation}
\label{defdensityN}
\hat{\rho}(\vec{x})\equiv \frac{1}{N}\sum_{i=1}^N \delta({\vec{x}-\vec{x}_i}),
\end{equation}
which is decomposed as
\beq
\hat{\rho}(\vec{x})=\hat{\rho}^{\uparrow}(\vec{x})
+\hat{\rho}^{\downarrow}(\vec{x}),
\eeq
with
\beq
\hat{\rho}^{\uparrow}(\vec{x})\equiv \frac{1}{N}
\sum_{i=1}^N \delta({\vec{x}-\vec{x}_i})\ket{\uparrow}_i\bra{\uparrow}_i
\eeq
and 
\beq
\hat{\rho}^{\downarrow}(\vec{x})\equiv \frac{1}{N}\sum_{i=1}^N 
\delta({\vec{x}-\vec{x}_i})\ket{\downarrow}_i\bra{\downarrow}_i.
\eeq

\end{document}